\documentclass[aps,a4paper,amsmath,amssymb,twocolumn,
floatfix,superscriptaddress]{revtex4}
\usepackage{bm,amsmath,amssymb,amsfonts}
\usepackage[dvips]{graphicx}
\usepackage[utf8]{inputenc}
\usepackage[normalem]{ulem}
\usepackage{xcolor}
\definecolor{dblue}{rgb}{0.0,0.0,0.7}
\definecolor{dred}{rgb}{0.9,0.0,0.0}
\definecolor{dgreen}{rgb}{0,0.8,0.0}
\usepackage{hyperref}
\usepackage{cancel}
\hypersetup{
    pdfnewwindow=true,      
    colorlinks=true,       
    linkcolor=red,          
    citecolor=blue,        
    filecolor=blue,      
    urlcolor=blue        
}
\newcommand{\ie}{{\it i.\,e.}}

\newcommand{\etal}{{\it et al.}}
\newcommand{\viz}{{\it viz.}} 

\begin{document}
\title{Engineering flux-controlled flat bands and topological states in a Stagome lattice}
\author{Biplab Pal}
\thanks{Corresponding author}
\email[E-mail: ]{biplab@nagalanduniversity.ac.in}
\affiliation{Department of Physics, School of Sciences, 
Nagaland University, Lumami-798627, Nagaland, India}
\author{Georges Bouzerar}
\email[E-mail: ]{georges.bouzerar@neel.cnrs.fr}
\affiliation{Universit\'{e} Grenoble Alpes, CNRS, Institut NEEL, 
F-38042 Grenoble, France}
\date{\today}
\begin{abstract}
We present the Stagome lattice, a variant of the Kagome lattice, where one can make any of the bands 
completely flat by tuning an externally controllable magnetic flux. This systematically allows the 
energy of the flat band to coincide with the Fermi level. We have analytically calculated the compact 
localized states associated to each of these flat bands appearing at different values of the magnetic 
flux. We also show that, this model features nontrivial topological properties with distinct integer 
values of the Chern numbers as a function of the magnetic flux. We argue that this mechanism for 
making any of the bands exactly flat could be of interest to address the flat-band superconductivity 
in such a system. Additionally, we show that our results are robust even in the presence of a small 
amount of disorder. Furthermore, we believe that the phenomenon of photonic flat band localization 
could be studied in the Stagome lattice structure, designed for instance using femtosecond laser 
induced single-mode waveguide arrays.
\end{abstract}
\maketitle
\section{Introduction}
\label{sec:Intro}
The study of flat band (FB) physics has come up as one of most intriguing field 
of research in condensed matter physics and material science in recent 
years~\cite{FB-Review-Flach,FB-Review-Chen,FB-Review-Vicencio}.
FBs offer highly degenerate manifold of single particle states and provide 
an ideal platform to study a wide range of strongly correlated phenomena, such as 
ferromagnetism~\cite{FB-ferromagnetism-1,FB-ferromagnetism-2,FB-ferromagnetism-3,
FB-ferromagnetism-Georges}, superconductivity~\cite{FB-superconductivity-1,
FB-superconductivity-2,FB-superconductivity-3}, quantum Hall 
effect~\cite{Wen-PRL-2011,Das-Sarma-PRL-2011,Neupert-PRL-2011}, 
and inverse Anderson transition~\cite{FB-Anderson-transition-1,
FB-Anderson-transition-2}, to name a few of them. In tight-binding lattice 
models, FB states originate from destructive quantum interference of electron 
hoppings between neighboring sites. 
At the FB energies, one can often construct 
exact real space eigenstates which have strictly zero support outside a region 
of the lattice that spans over a few unit cells. Such eigenstates are known as 
compact localized states (CLS)~\cite{Ajith-PRB-2017,Pal-Saha-PRB-2018,
Pal-PRB-2018,Bhatta-Pal-PRB-2019,Singular-FB-Rhim-2021}. 
One may view these CLS to be acting like 
a prison for the electrons at the FB energy, from which they cannot escape. 
The appearance of FBs in simple tight-binding 2D-lattice models also often 
allows us to bring out interesting topological properties~\cite{Wen-PRL-2011,
Das-Sarma-PRL-2011,Neupert-PRL-2011,Pal-PRB-2018,Bhatta-Pal-PRB-2019}. 

FB models are not just theoretical artifacts anymore, in recent times they have 
received a strong attention in photonic lattice experiments~\cite{Vicencio-PRL-2015,
Mukherjee-PRL-2015,Mukherjee-OL-2015,Longhi-OL-2014,Xia-OL-2016,Zong-OE-2016,
Weimann-OL-2016,Chen-Suepr-Honeycomb-AOM-2020,Chen-Super-Kagome-OL-2023,
Denz-Extended-Lieb-FB,Hahafi-fractal-FB,Chen-fractal-FB-1,Chen-fractal-FB-2}, 
where the lattice is composed of evanescently coupled two-dimensional arrays of 
single-mode optical waveguides, fabricated using very sophisticated femtosecond 
laser writing techniques~\cite{Vicencio-PRL-2015, Mukherjee-PRL-2015,Mukherjee-OL-2015}. 
In addition, over the last few years, the field of FB has evolved significantly with 
the experimental identification of electronic FBs in moir\'{e} superlattice 
structures~\cite{Balents-Moire-FB-1,LeRoy-Moire-FB-2,Li-Moire-FB-3}, where the FBs 
appear for certain magic angles between the twisted layers. Apart from these, 
interests in FB research has also expanded its wings in the field of chemical 
sciences, where people have realized FBs in metal-organic materials based on 
density functional theory (DFT) driven calculations~\cite{chen-jmca18,ni-nanoscale18,
su-apl18,jiang-nanoscale19,zhang-prb19,jiang-natcomm19,Reganult-Nat-2022}. A very recent 
interesting development in the field of FB research is the realization of FBs based 
on topo-electrical circuits~\cite{Zhang-EC-FB-1,Zhang-EC-FB-2,Zhang-EC-FB-3,Flach-EC-FB} 
and also in circuit Quantum Electrodynamics (QED) lattice models~\cite{Circuit-QED-FB}.
\begin{figure}[ht]
\centering
\includegraphics[clip, width=0.9\columnwidth]{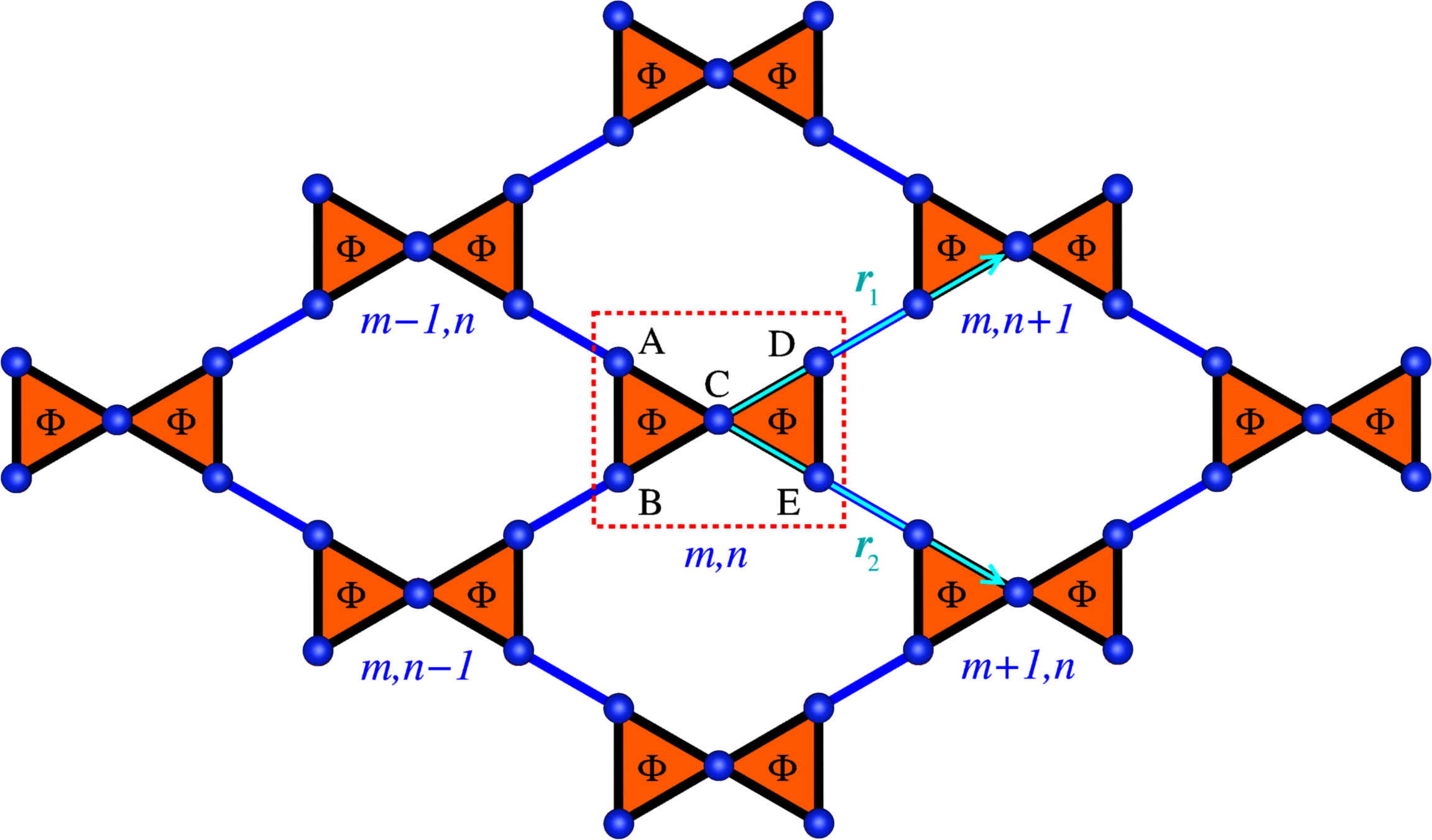}
\caption{Schematic diagram of a \textit{Stagome} lattice geometry. The unit cell 
of the lattice is marked by the red dotted box. Both the triangular plaquettes 
in each cell are pierced by an uniform external magnetic flux $\Phi$ directed 
in the anti-clockwise direction.}
\label{fig:Lattice-model}
\end{figure}
%
\begin{figure}[ht]
\centering
\includegraphics[clip, width=0.9\columnwidth]{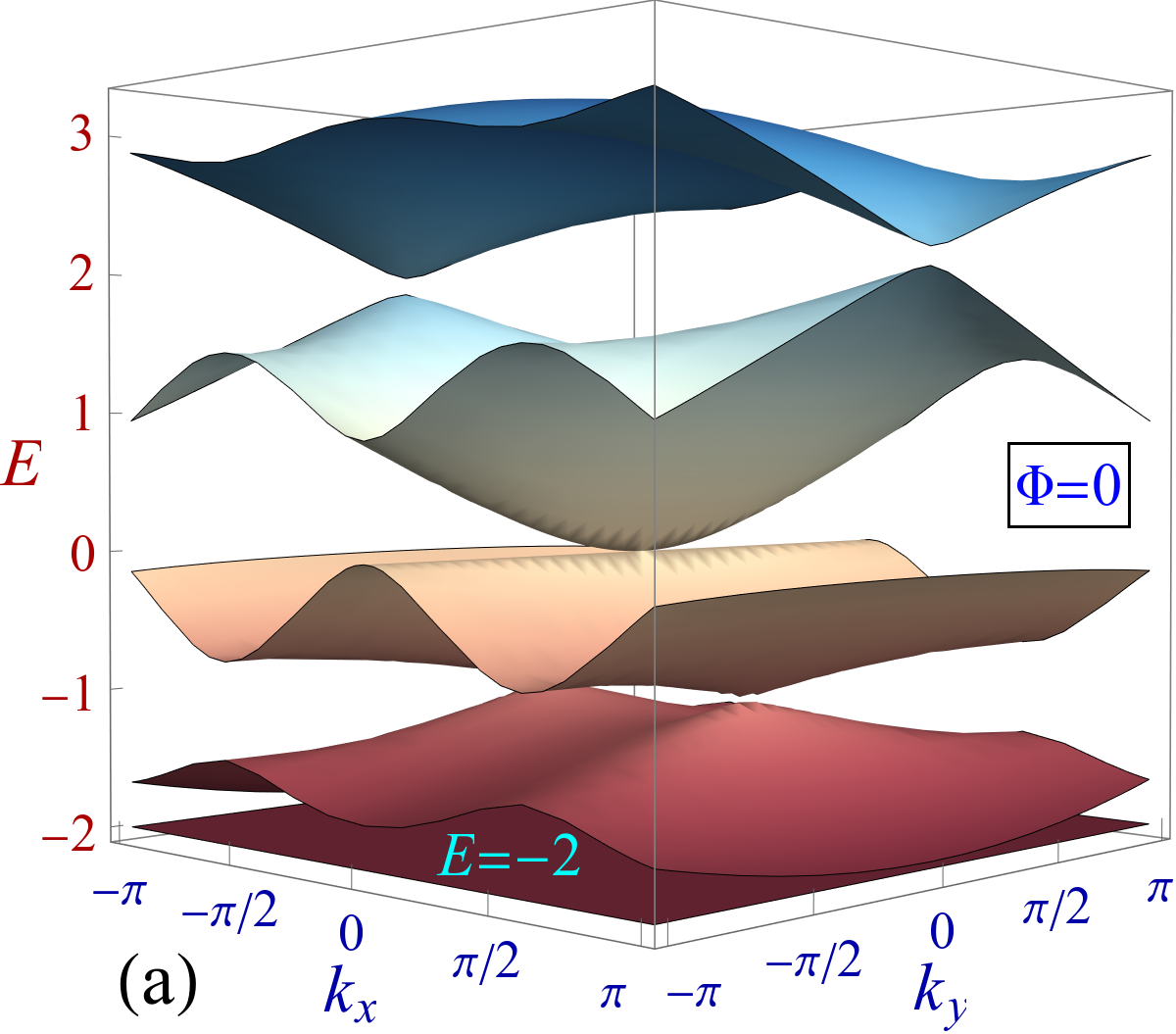}
\vskip 0.4cm
\includegraphics[clip, width=0.9\columnwidth]{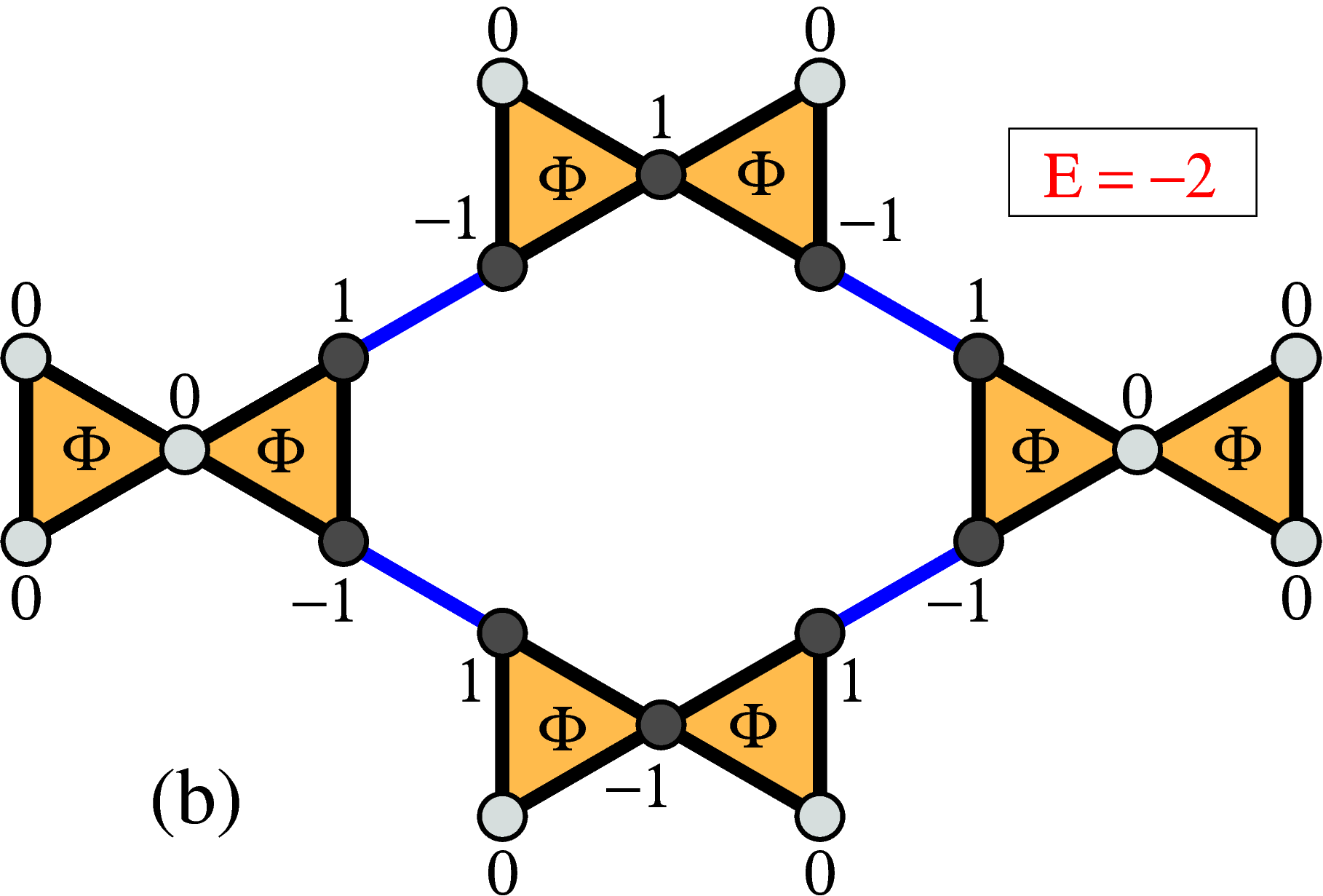} 
\caption{(a) Electronic band structure for the Stagome lattice in absence of 
magnetic flux (\ie, $\Phi=0$) where the bottom band is completely 
flat ($E=-2$), and (b) wave function amplitude distribution of the compact 
localized state corresponding to the FB.}
\label{fig:Band-n-CLS_without-flux}
\end{figure}
%

FB systems have many practical technological applications, particularly in the 
realm of photonics, in addition to their inherent interests in fundamental physics. 
This includes diffraction-free propagation of light~\cite{Vicencio-PRL-2015,
Mukherjee-PRL-2015}, telecommunication and sensing~\cite{Vicencio-PRL-2015}, 
distortion-free image processing~\cite{Mukherjee-PRL-2015,Xia-OL-2016,Zong-OE-2016}, 
generating slow light~\cite{Baba-Nat-Photon-2008,Nandy-PRA-2016}, and even 
FB lasing mechanism~\cite{Longhi-OL-2019} among others. This is due to the 
fact, that nonlinear effects in systems with nondispersive bands are relatively easy 
to implement experimentally, especially for investigations involving optics. 
In addition to this, FB models are also realized experimentally using ultracold 
atoms in optical lattices~\cite{takahashi-sciadv15,takahashi-prl17} and 
exciton-polariton condensates~\cite{masumoto-njp12,baboux-prl16}, which offers 
a novel platform for studying coherent phases of light and matter. People have 
proposed and explored a broad range of intriguing lattice models over time, 
including the Kagome lattice~\cite{Wen-PRL-2011,Neupert-PRL-2011}, 
Lieb lattice~\cite{Vicencio-PRL-2015,Mukherjee-PRL-2015}, 
diamond chains~\cite{Mukherjee-OL-2015,Longhi-OL-2014}, 
checkerboard lattice~\cite{Das-Sarma-PRL-2011}, sawtooth chain~\cite{Weimann-OL-2016}, 
and so on, in order to recognize these possible applications appearing in FB systems. 
More recently, researchers have expanded the horizon by proposing and studying more 
exotic structures like diamond-octagon lattice~\cite{Pal-PRB-2018,FB-superconductivity-3}, 
extended Lieb lattice~\cite{Bhatta-Pal-PRB-2019,Denz-Extended-Lieb-FB}, 
super-honeycomb lattice~\cite{Chen-Suepr-Honeycomb-AOM-2020}, 
super-Kagome lattice~\cite{Chen-Super-Kagome-OL-2023}, and even fractal-like 
geometries~\cite{Pal-Saha-PRB-2018,Hahafi-fractal-FB,Chen-fractal-FB-1,
Chen-fractal-FB-2,Biswas-fractal-FB}. In fact, one can find out a comprehensive 
list of various possible FB models and their material realizations 
in Refs.~\cite{FB-catalog-npj-Comp-Mat,FB-classification-Calugaru}. 

To the best of our knowledge, there have been no instances where one could make any 
of the all possible bands appearing in the system to be flat at a will. The possibility 
of controlling the position of the FB and adjusting it with the Fermi energy could open 
interesting avenues from a technological point of view. 
Here, we address this issue by proposing a new two-dimensional lattice with five-site 
unit cell, as depicted  in Fig.~\ref{fig:Lattice-model}, which is known as 
the \emph{Stagome} lattice. We show that for this particular geometry we, indeed, 
can realize this phenomena. For this model, we can precisely make any of the bands to 
be completely flat by controlling the value of an externally tunable magnetic flux 
(see Fig.~\ref{fig:Lattice-model}). 
We remark that, in the literature, in recent times a real chemically synthesized 
Stagome compound has been reported~\cite{Stagome-expt-synthesis}, 
and its magnetic susceptibility and heat capacity measurements were done. However, we 
would like to clarify that our proposed model is much more simpler and not a descriptive 
of the real Stagome compound. For instance, we do not consider the spin degree of freedom 
in our model which is pertinent in a real Stagome compound~\cite{Stagome-expt-synthesis}. 
We consider the underlying lattice geometry of the real Stagome compound and study certain 
interesting aspects of the flat band physics in a simplified Stagome lattice model. 
The name ``Stagome" is used because its geometry is intermediate between a Star lattice 
and a Kagome lattice. This is adopted from Ref.~\cite{Stagome-expt-synthesis}. 

In this work, we show that one can tune the position of the flat band at 
different energies in a flexible way by controlling the magnetic flux in the system. This 
gives us an advantage to set the Fermi level at different FB energies and study the Physics 
of the Stagome lattice at different filling factors. We have explored the robustness of 
these FBs when we go beyond a perfectly clean system, such as the effect of the disorder and 
long-range hopping. We have carried out a detailed analysis of the CLS corresponding to the 
FBs in this system. The CLS distribution over a relatively large number of sites indicates 
that this lattice model could be useful for the study of FB superconductivity in future. 

The manuscript is arranged as follows: In Sec.~\ref{sec:model}, we introduce the Hamiltonian 
and show our results in the case of zero flux in Sec.~\ref{sec:FB-without-flux}. 
Then, in Sec.~\ref{sec:FB-with-flux}, we show that one can make other bands become 
completely flat in a systematic way by tuning the magnetic flux. In all cases, we 
construct the CLS and compute the flux dependent density of states (DOS). 
This is followed by Sec.~\ref{sec:topo-properties}, where we present the topological 
properties of the model by calculating the Berry curvature and the Chern 
numbers~\cite{Wen-PRL-2011,Das-Sarma-PRL-2011,Neupert-PRL-2011,Pal-PRB-2018} in the 
presence of the magnetic flux. Then, in Sec.~\ref{sec:disorder}, we address the impact 
of the disorder on the features observed in the clean system. 
Finally, we make our conclusion in Sec.~\ref{sec:conclu}.
%
\begin{figure*}[ht]
\centering
\includegraphics[clip, width=0.8\columnwidth]{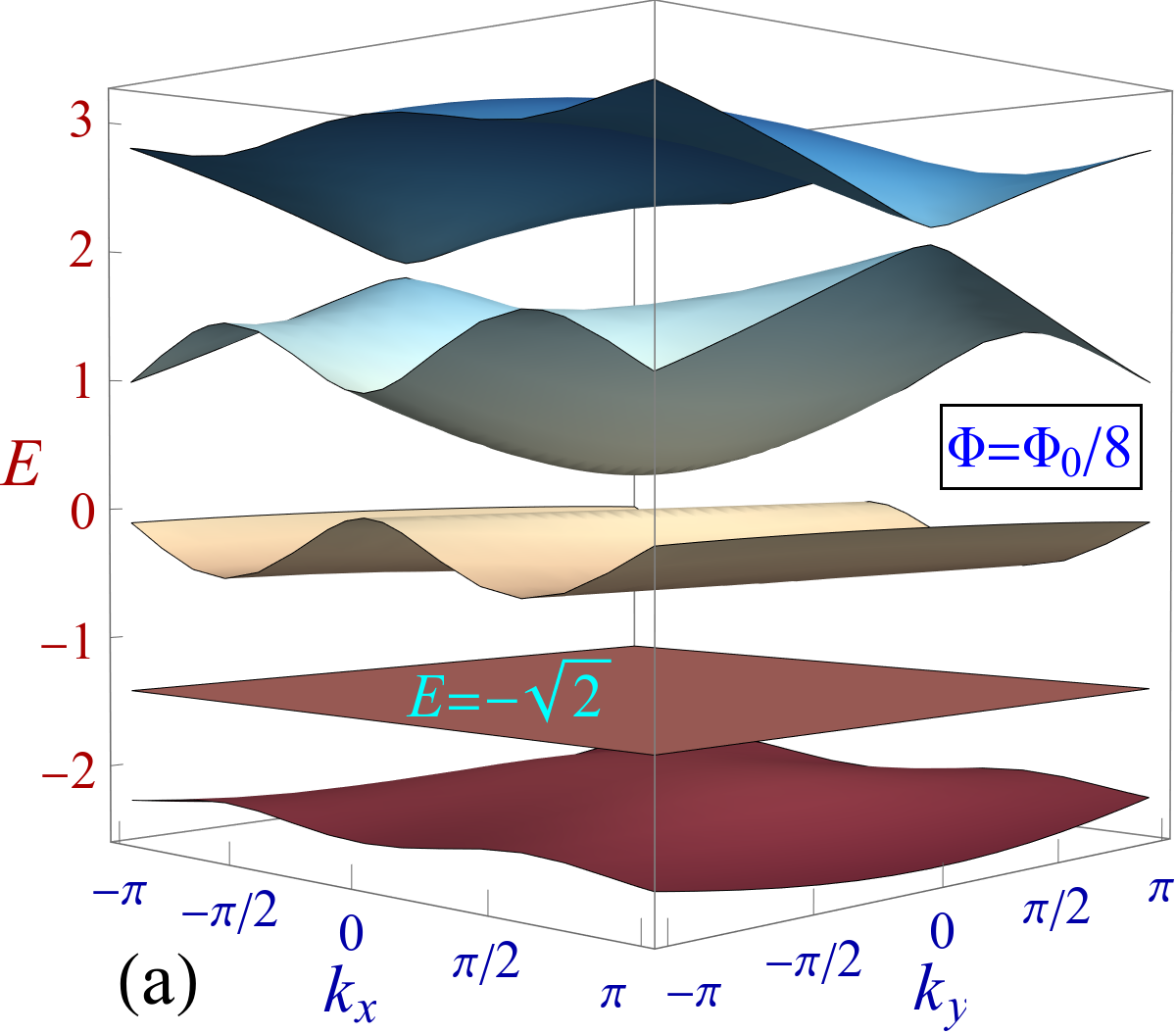}
\quad
\includegraphics[clip, width=0.8\columnwidth]{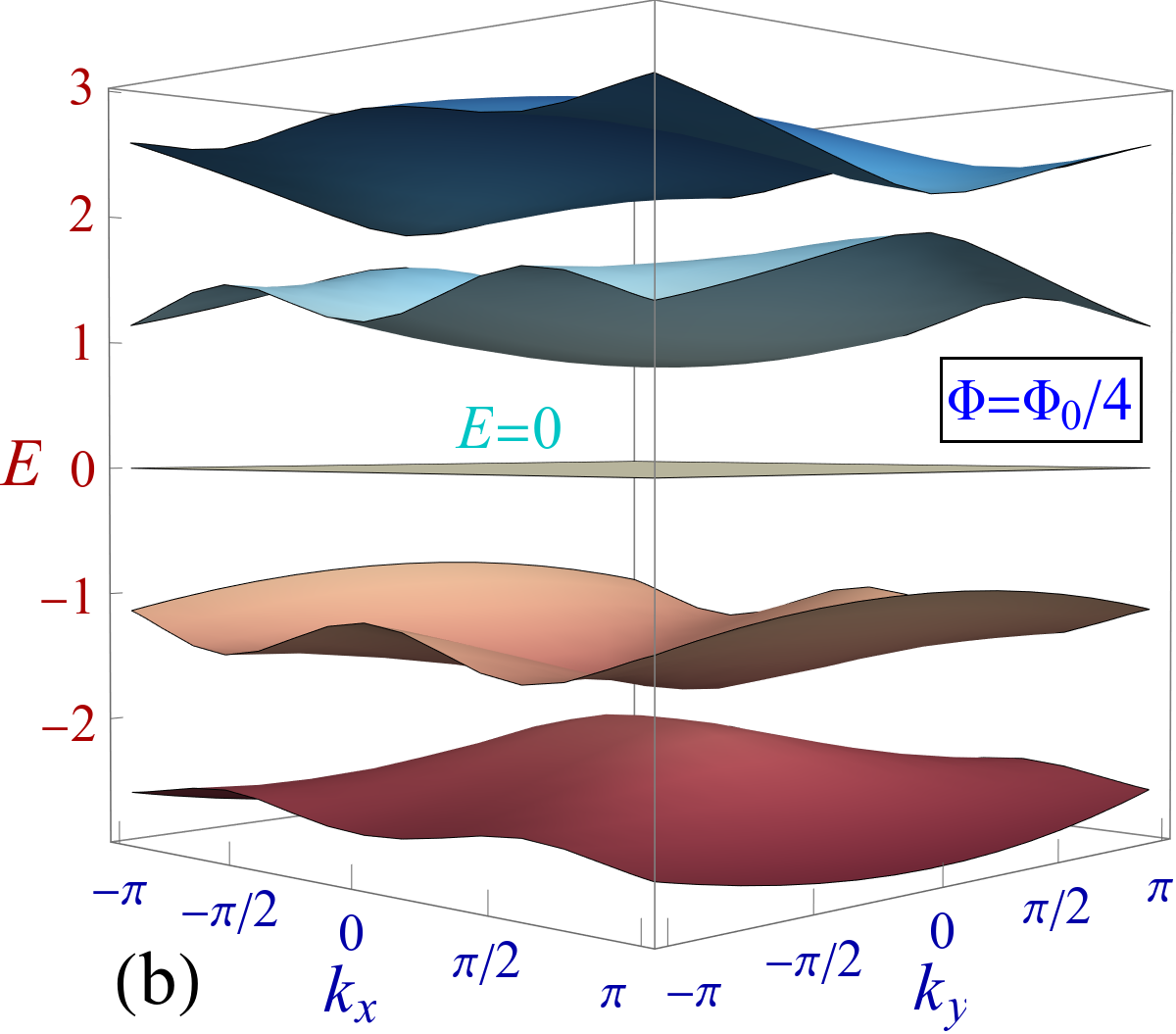}
\vskip 0.4cm
\includegraphics[clip, width=0.8\columnwidth]{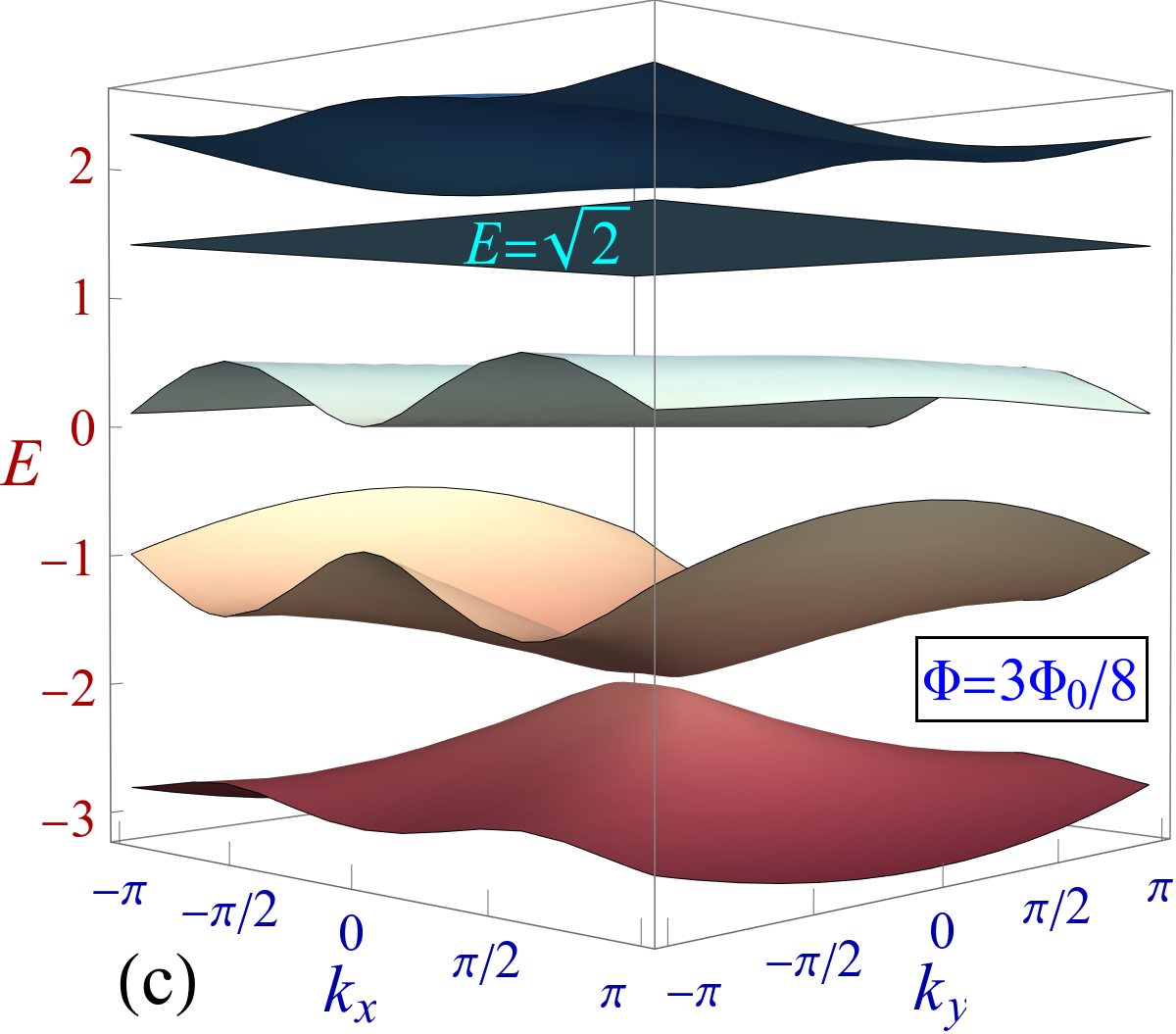}
\quad
\includegraphics[clip, width=0.8\columnwidth]{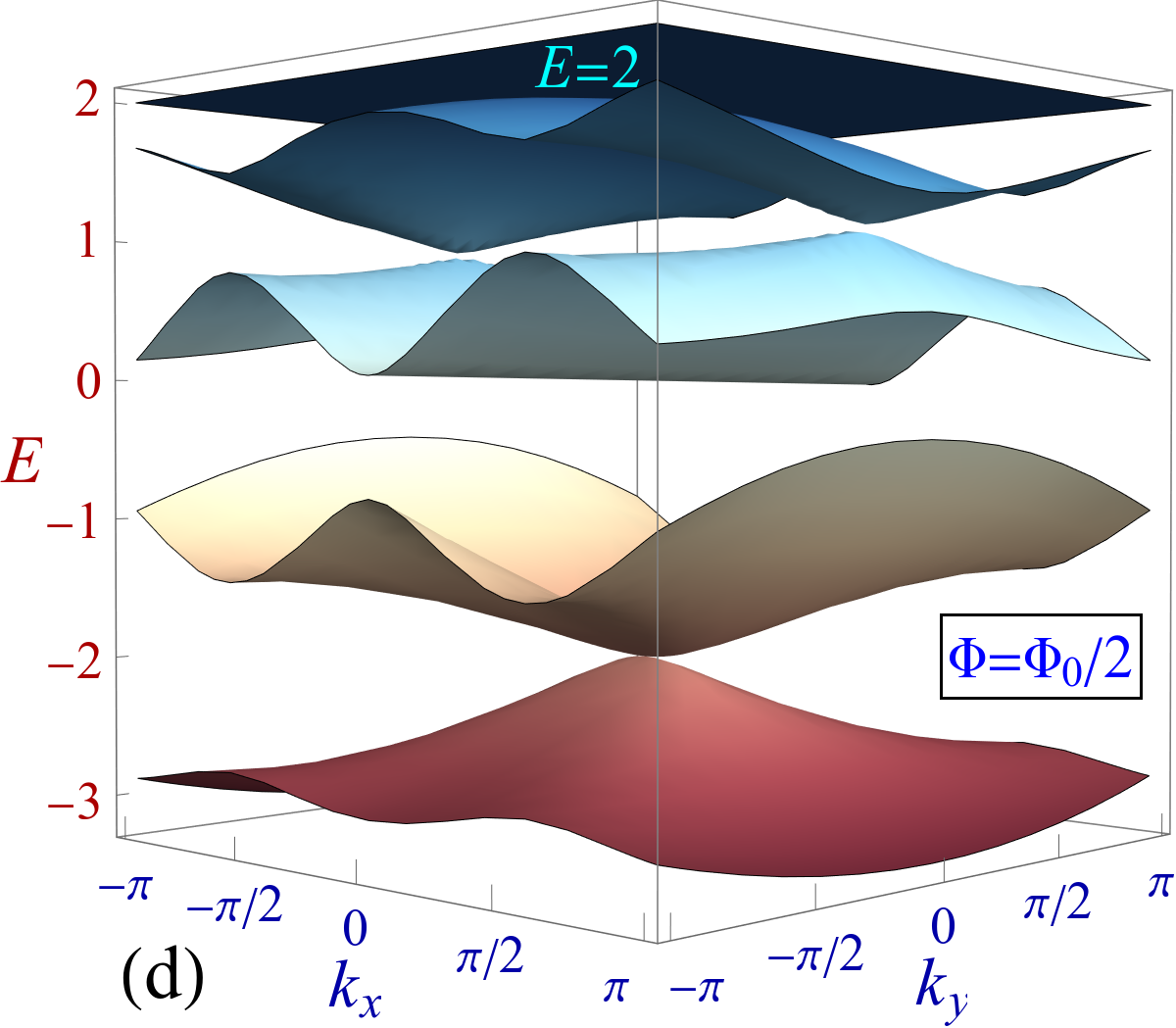}
\caption{Electronic band structure for the Stagome lattice for four different 
special values of the magnetic flux, \viz, 
(a) $\Phi=\Phi_{0}/8$ where the second band is completely flat ($E=-\sqrt{2}$), 
(b) $\Phi=\Phi_{0}/4$ where the third band is the FB ($E=0$), 
(c) $\Phi=3\Phi_{0}/8$ where the fourth band is the FB ($E=\sqrt{2}$), 
and (d) $\Phi=\Phi_{0}/2$ where the top band is completely flat ($E=2$).}
\label{fig:Band-structures_with-flux}
\end{figure*}
%
\begin{figure*}[ht]
\centering
\includegraphics[clip, width=0.8\columnwidth]{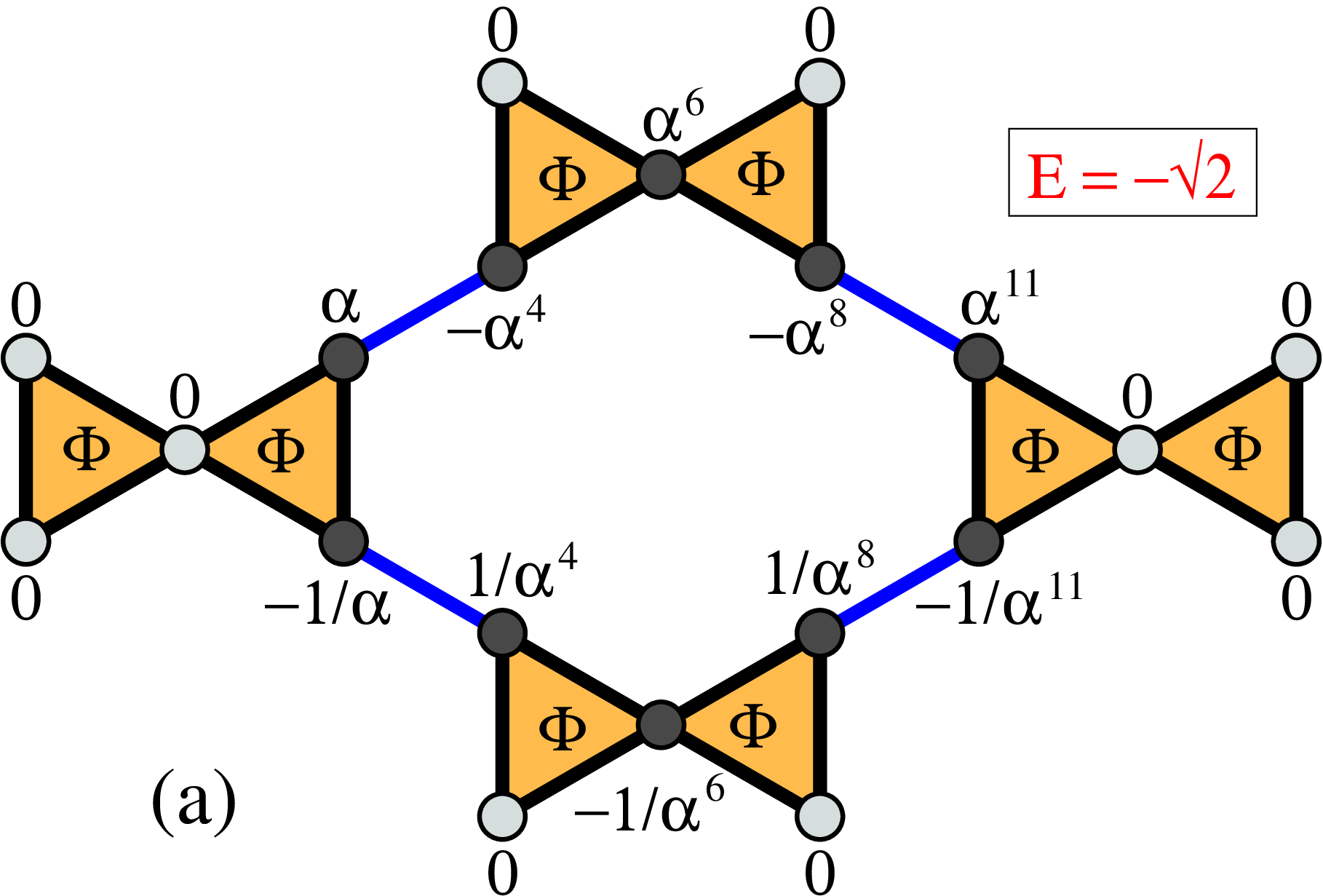}
\quad
\includegraphics[clip, width=0.8\columnwidth]{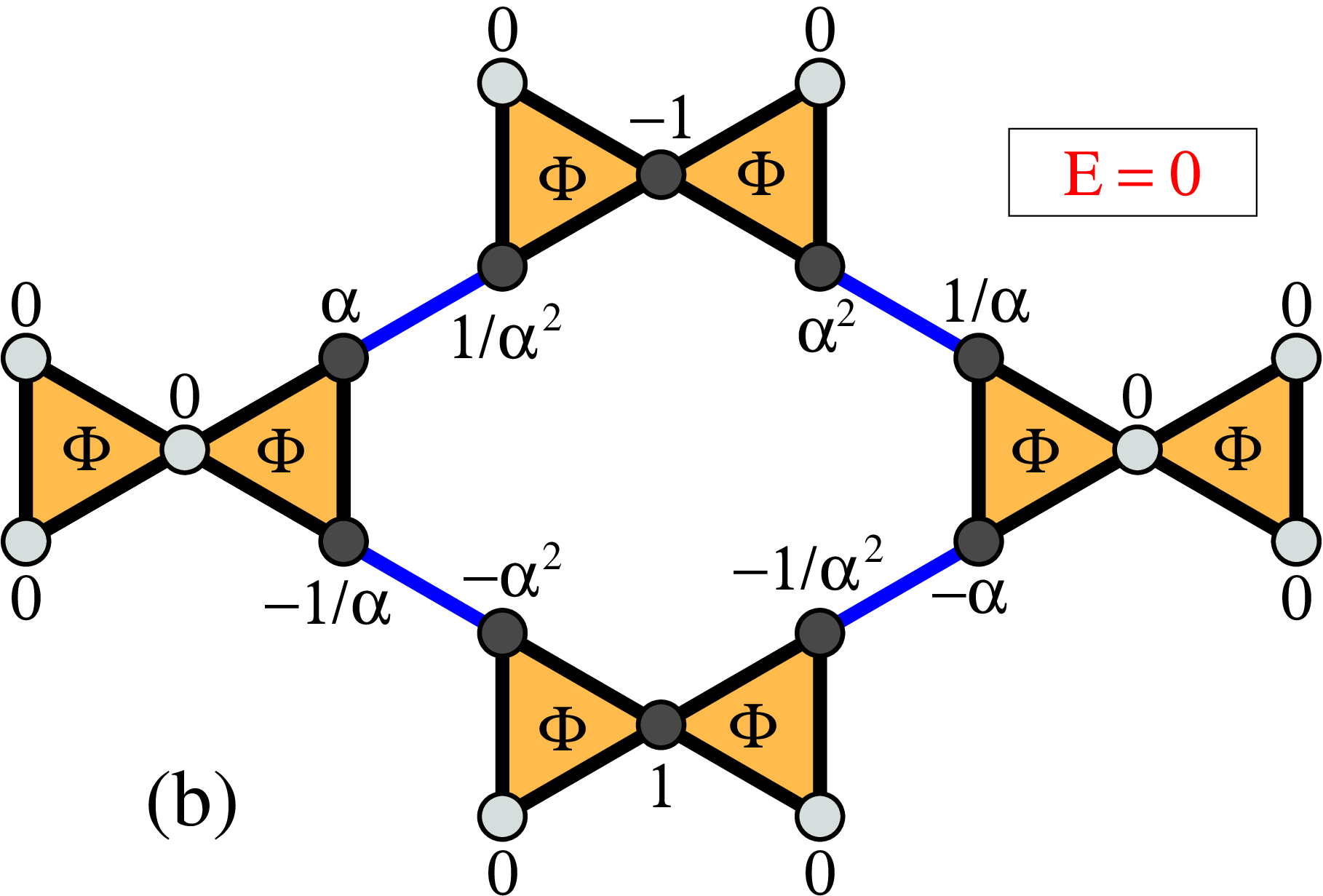}
\vskip 0.4cm
\includegraphics[clip, width=0.8\columnwidth]{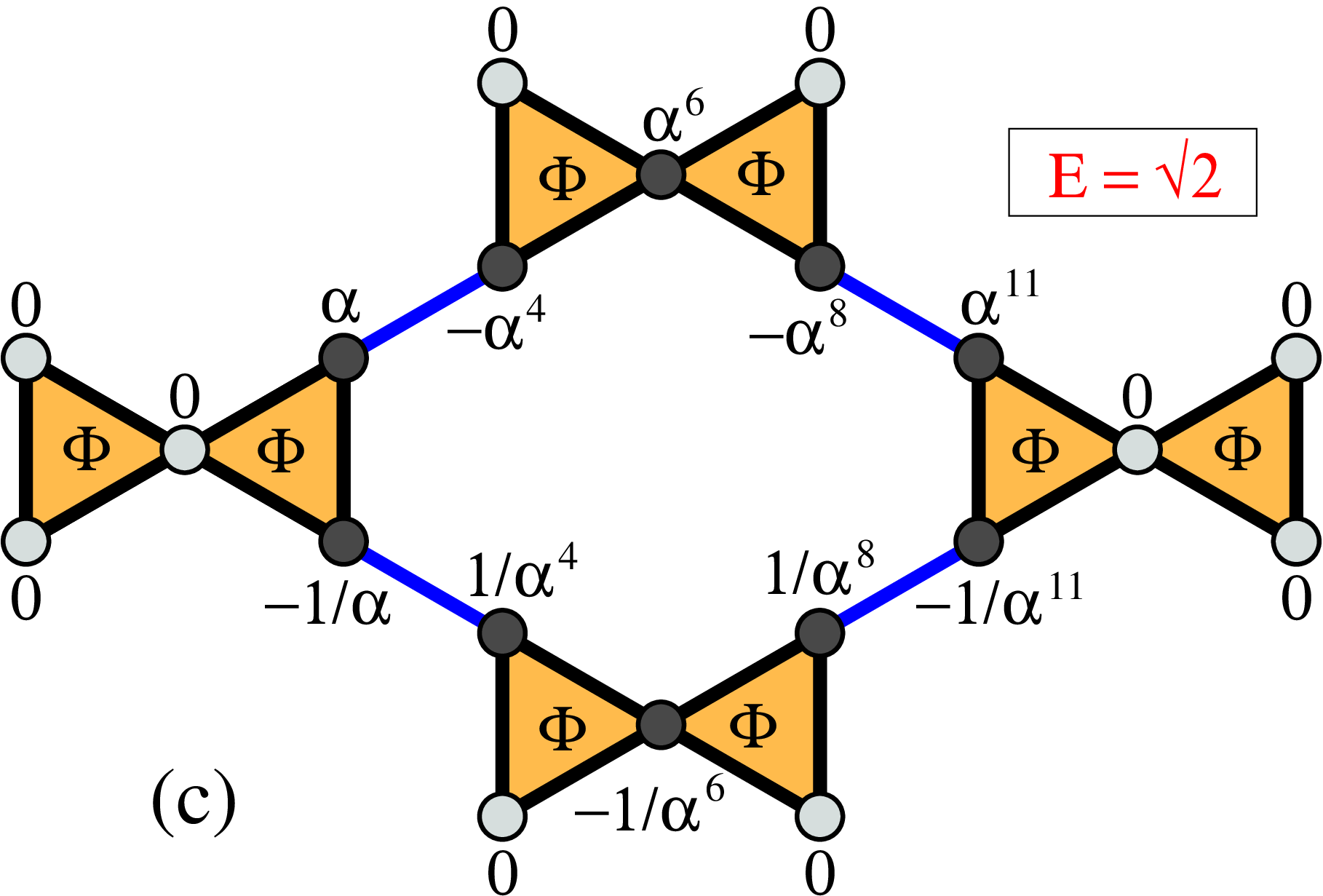}
\quad
\includegraphics[clip, width=0.8\columnwidth]{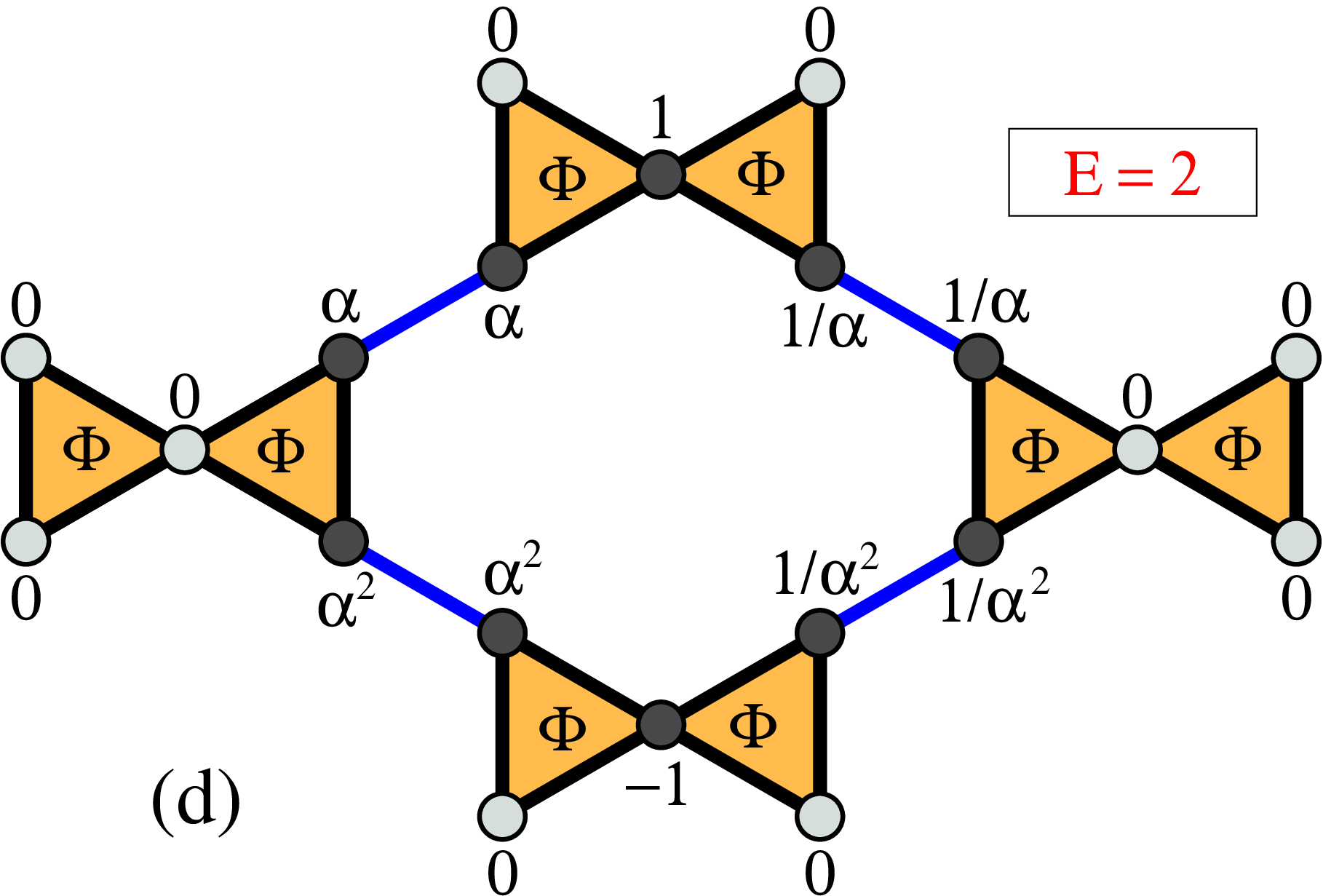}
\caption{The distribution of the wave function amplitudes for the compact localized 
states corresponding to the FBs appearing in Fig.~\ref{fig:Band-structures_with-flux} 
(a)-(d). The light grey circles represent sites with zero amplitudes and the black 
circles represent sites with nonzero amplitudes. Here $\alpha = e^{i\theta}$, 
with $\theta = 2\pi\Phi / 3\Phi_{0}$.}
\label{fig:CLS_with-flux}
\end{figure*}
\section{The lattice model}
\label{sec:model} 
In real space the tight-binding 
Hamiltonian of the Stagome lattice reads, 
\begin{widetext}
\begin{equation}
\bm{H} = \sum_{m,n}\bigg[
\underbrace{ \bigg(\sum_{\alpha}\epsilon_{\alpha} 
c_{m,n,\alpha}^{\dagger}c_{m,n,\alpha}\bigg) 
}_\text{Onsite energy terms}
+ 
\underbrace{ \bigg(\sum_{\langle \alpha,\beta \rangle}t_{\alpha,\beta} 
c_{m,n,\alpha}^{\dagger}c_{m,n,\beta} + \textrm{h.c.}\bigg) 
}_\text{Intra-cell hopping terms}
+ 
\underbrace{ \bigg(\sum_{\langle \alpha,\beta \rangle}\lambda_{\alpha,\beta} 
\big(c_{m,n,\alpha}^{\dagger}c_{m+1,n,\beta} + c_{m,n,\alpha}^{\dagger}c_{m,n+1,\beta}\big) + \textrm{h.c.}\bigg) 
}_\text{Inter-cell hopping terms}
\bigg],
\label{eq:hamilton-wannier}
\end{equation}
\end{widetext} 
where $(m,n)$ indices stand for the cell index as shown in Fig.~\ref{fig:Lattice-model}, 
$\alpha$ is the atomic site index within a cell, and $\alpha \in \{A,B,C,D,E\}$. 
$\langle \alpha,\beta \rangle$ indicates the nearest neighbor pairs. 
$c_{m,n,\alpha}^{\dagger}(c_{m,n,\alpha})$ represents the creation (annihilation) 
operator for an electron at site $\alpha$ in the $(m,n)$-th cell. 
Here, we ignore the spin degree of freedom which is irrelevant. $\epsilon_{\alpha}$ 
denotes the onsite energy, $t_{\alpha,\beta}$ (resp. $\lambda_{\alpha,\beta}$) is 
the intra-cell (resp. inter-cell) nearest neighbor hopping integral for orbitals 
in the same (resp. different) cell. For simplicity, we choose $\epsilon_{\alpha} = 0$ for 
all $\alpha$-sites, $t_{\alpha,\beta} = t$ and $\lambda_{\alpha,\beta} = \lambda$. 
We first focus on the case of a clean system and will discuss the impact 
of the disorder in the last section. 

We introduce an uniform external magnetic flux $\Phi$ piercing through the 
triangular plaquettes in each cell as illustrated in Fig.~\ref{fig:Lattice-model}. 
Hence, along the closed triangular plaquettes, the hopping on 
each of the sides of the triangle picks up a phase factor 
$\theta = 2\pi\Phi/3\Phi_{0}$, where $\Phi_{0}=2\pi\hbar/e$ ~\cite{A-B-Phase}. 
Thus, $t \rightarrow t e^{\pm i\theta}$, where the sign $\pm$ indicates the direction 
of the hopping. To get the band structure of our model, we perform a standard Fourier 
transform of Eq.~\eqref{eq:hamilton-wannier} which leads to,
\begin{equation}
\bm{H} = \sum_{\bm{\vec{k}}} \bm{C}^{\dagger}_{\bm{\vec{k}}} 
\bm{\mathcal{H}}(\bm{\vec{k}})\bm{C}_{\bm{\vec{k}}},
\label{eq:hamilton-k-space} 
\end{equation}
where the $5\times5$ matrix $\bm{\mathcal{H}}(\bm{\vec{k}})$ is given by, 
\begin{align}
\bm{\mathcal{H}}(\bm{\vec{k}}) =
\left(\def\arraystretch{1.5} \begin{matrix}
0 &  te^{i\theta}  &  te^{-i\theta}  &  0  
&  \lambda e^{-i\bm{\vec{k}}\cdot\bm{\vec{r}_{2}}} \\
te^{-i\theta}  &  0  &  te^{i\theta}  &  
\lambda e^{i\bm{\vec{k}}\cdot\bm{\vec{r}_{1}}}  &  0 \\
te^{i\theta}  &  te^{-i\theta}  &  0 &  
te^{-i\theta}  &  te^{i\theta}  \\
0  &  \lambda e^{-i\bm{\vec{k}}\cdot\bm{\vec{r}_{1}}}  &  
te^{i\theta}  &  0  &  te^{-i\theta}  \\
\lambda e^{i\bm{\vec{k}}\cdot\bm{\vec{r}_{2}}}  &  
0  &  te^{-i\theta}  &  te^{i\theta}  &  0  \\
\end{matrix}\right)
\label{eq:hamilton-bloch}
\end{align}
and the spinor is defined as,
\begin{align*} 
\bm{C}^{\dagger}_{\bm{\vec{k}}} = 
\left(\begin{matrix}
c^{\dagger}_{\bm{\vec{k}},A}  &  c^{\dagger}_{\bm{\vec{k}},B}  
&  c^{\dagger}_{\bm{\vec{k}},C} & c^{\dagger}_{\bm{\vec{k}},D} 
& c^{\dagger}_{\bm{\vec{k}},E}
\end{matrix}\right),
\end{align*} 
and,
$\bm{\vec{k}}\cdot\bm{\vec{r}}_{1} = 
\dfrac{\sqrt{3}k_{x}}{2} + \dfrac{k_{y}}{2}$, 
$\bm{\vec{k}}\cdot\bm{\vec{r}}_{2} = 
\dfrac{\sqrt{3}k_{x}}{2} - \dfrac{k_{y}}{2}$ and the lattice constant $a$ is set to $1$.

Let us briefly comment about the possibility to achieve such a flux distribution 
experimentally since it clearly cannot be obtained by simply applying a uniform 
magnetic field. Recent advances and progress in the field of cold atoms have made 
it possible to achieve effective staggered fields or fluxes using a technique that 
relies on atom tunneling assisted by Raman transitions~\cite{aidelsburger,jaksch,gerbier,ashvin}.

In what follows, we diagonalize $\bm{\mathcal{H}}(\bm{\vec{k}})$ to elicit 
the exotic FB physics and topological properties of the Stagome lattice. We set the 
intra-cell and inter-cell hopping parameters to be equal to each other, \ie, $\lambda=t=1$. 
We discuss the effect of inclusion of the next-nearest neighbor hopping 
in Appendix~\ref{appendix-1}. 
\section{Flat bands and CLS}
\subsection{Flat band and CLS in the absence of magnetic flux}
\label{sec:FB-without-flux} 
FBs arise from the strong localization of electronic wave functions caused by 
both destructive quantum interference and the local lattice 
topology ~\cite{Sutherland-PRB-1986}. In this section, we first discuss the 
case of the absence of the magnetic flux, \ie, $\Phi=0$. After diagonalization 
of the Hamiltonian in Eq.~\eqref{eq:hamilton-bloch}, we obtain a true FB at the 
bottom of the spectrum with energy $E_{\emph{FB}_{1}} = -2$ along with four 
other dispersive bands as depicted in Fig.~\ref{fig:Band-n-CLS_without-flux}(a). 
It is to note that, a FB at the bottom of the spectrum is apparently a common 
feature in the Kagome family, \viz, Kagome lattice~\cite{Zong-OE-2016}, 
super-Kagome lattice~\cite{Chen-Super-Kagome-OL-2023}, star lattice~\cite{chen-jpcm2012}, 
fractal-like lattices on a Kagome motif~\cite{Pal-Saha-PRB-2018} etc. 
However, in Stagome lattice, the other dispersive bands are not symmetric as 
compared to its counter parts in the Kagome family quoted above. In addition, 
a careful analysis reveals that (i) each band touches the next one and 
(ii) two Dirac cones appear at $E=-1$ and $E=2$. 
\begin{figure}[ht]
\centering
\includegraphics[clip, width=\columnwidth]{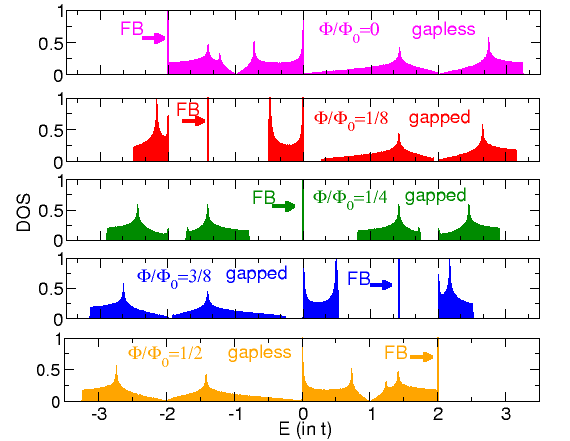}
\caption{The density of states (DOS) corresponding to a Stagome lattice for 
the special values of the external magnetic flux $\Phi_{p}$, for which a flat band 
appears in the spectrum. The position of the flat band and other features for 
the different values of the flux are indicated in the figure. Here, ``gapped" 
means the FB is well separated from the dispersive bands and ``gapless" means 
that there is band touching.}
\label{fig:DOS}
\end{figure}
%

We have constructed the CLS associated to the FB as depicted in 
Fig.~\ref{fig:Band-n-CLS_without-flux}(b). We remark that, the CLS spans 
over four unit cells which suggests large values of the quantum metric 
(QM) associated to the FB. The QM has been found to play a key role in the 
unconventional form of superconductivity that takes place in 
FBs~\cite{Peotta-Torma}. Hence, in the presence of attractive electron-electron 
interaction in the Stagome lattice, one may expect substantial values of 
the superfluid weight and thus potentially large superconducting critical 
temperatures~\cite{Kopnin,Heikkila,BoostQM-FB}. We recall that, the QM is 
related to the real part of the quantum geometric tensor (QGT) and provides 
a measure of the typical spread of the FB eigenstates~\cite{Marzari1,Marzari2}. 
On the other hand, the imaginary part of the QGT is the Berry curvature, 
which is discussed in Sec.~\ref{sec:topo-properties}.
\subsection{Flux-controlled flat band generation and CLS}
\label{sec:FB-with-flux}
Construction of FBs in presence of the magnetic flux (\ie, $\Phi \neq 0$) which 
breaks time reversal symmetry is a nontrivial task. In general, in this case, 
band gaps open up between the bands which is indeed observed in our model. 
We remark that, the presence of magnetic flux in each plaquette in the Kagome 
lattice destroys the FB~\cite{nagaosa-prb2000}. 
However, if the flux is staggered (\ie, some special combination of magnetic fluxes 
in two different plaquettes), one can regenerate a FB at a different energy 
in the Kagome lattice~\cite{green-prb2010}. Here, in the Stagome lattice, something 
interesting happens which was not observed previously in any other such lattice models. 
We show that, for a magnetic flux $\Phi = \Phi_{p} = p\times\frac{\Phi_{0}}{8}$, 
$p$ being integer, one can sequentially make each band in the band structure to be 
exactly flat as illustrated in Fig.~\ref{fig:Band-structures_with-flux}(a)-(d). 
\begin{figure}[ht]
\centering
\includegraphics[clip, width=\columnwidth]{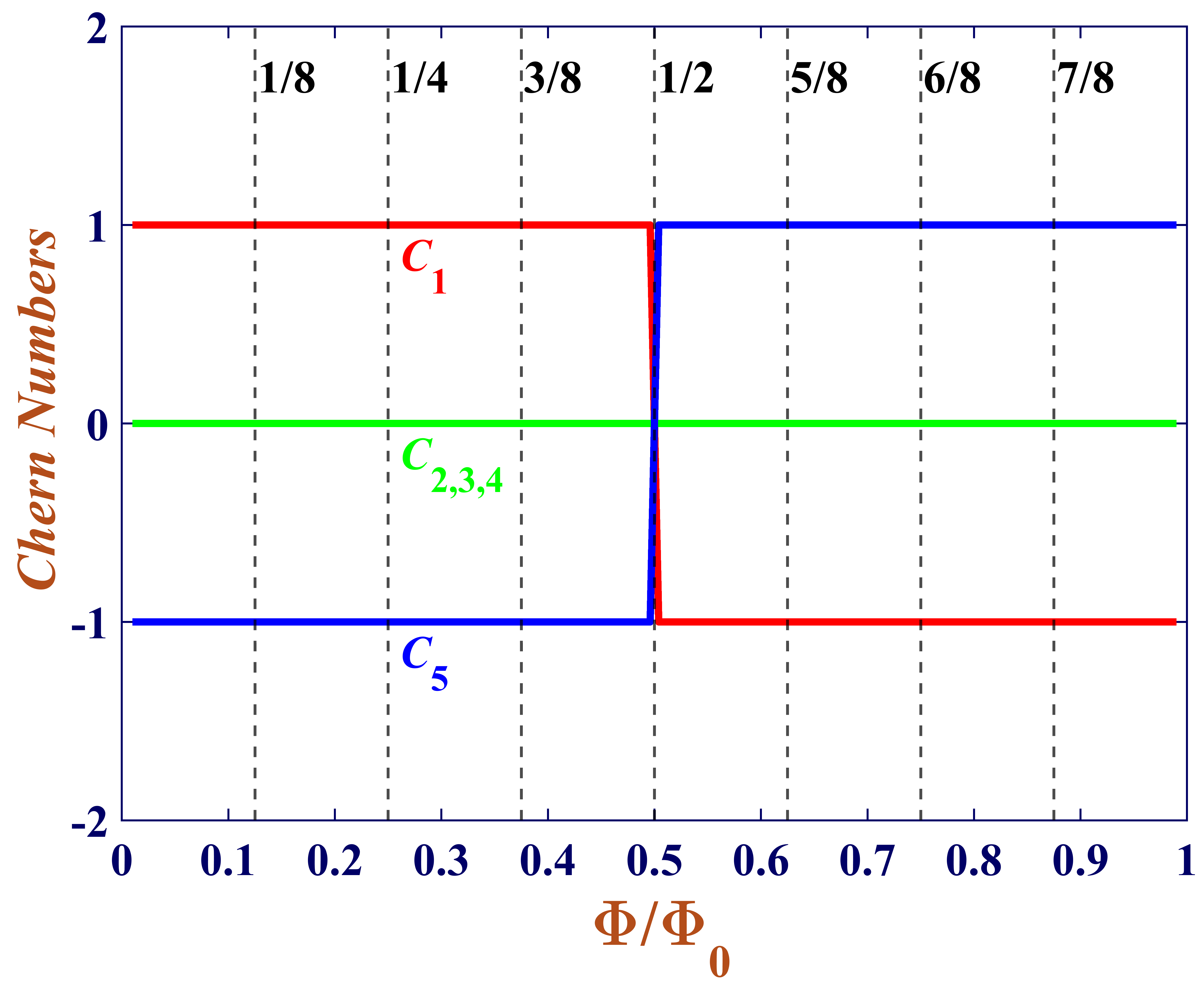}
\caption{The Chern numbers $\mathcal{C}_{n}$ as a function of 
the magnetic flux $\Phi$, the band index varies from $n=1$ to $5$. 
The vertical black dotted lines indicate the values of the magnetic flux, 
where there is a flat band appearing in the system.}
\label{fig:Chern-number}
\end{figure}
%
\begin{figure*}[ht]
\centering
\includegraphics[clip, width=0.55\columnwidth]{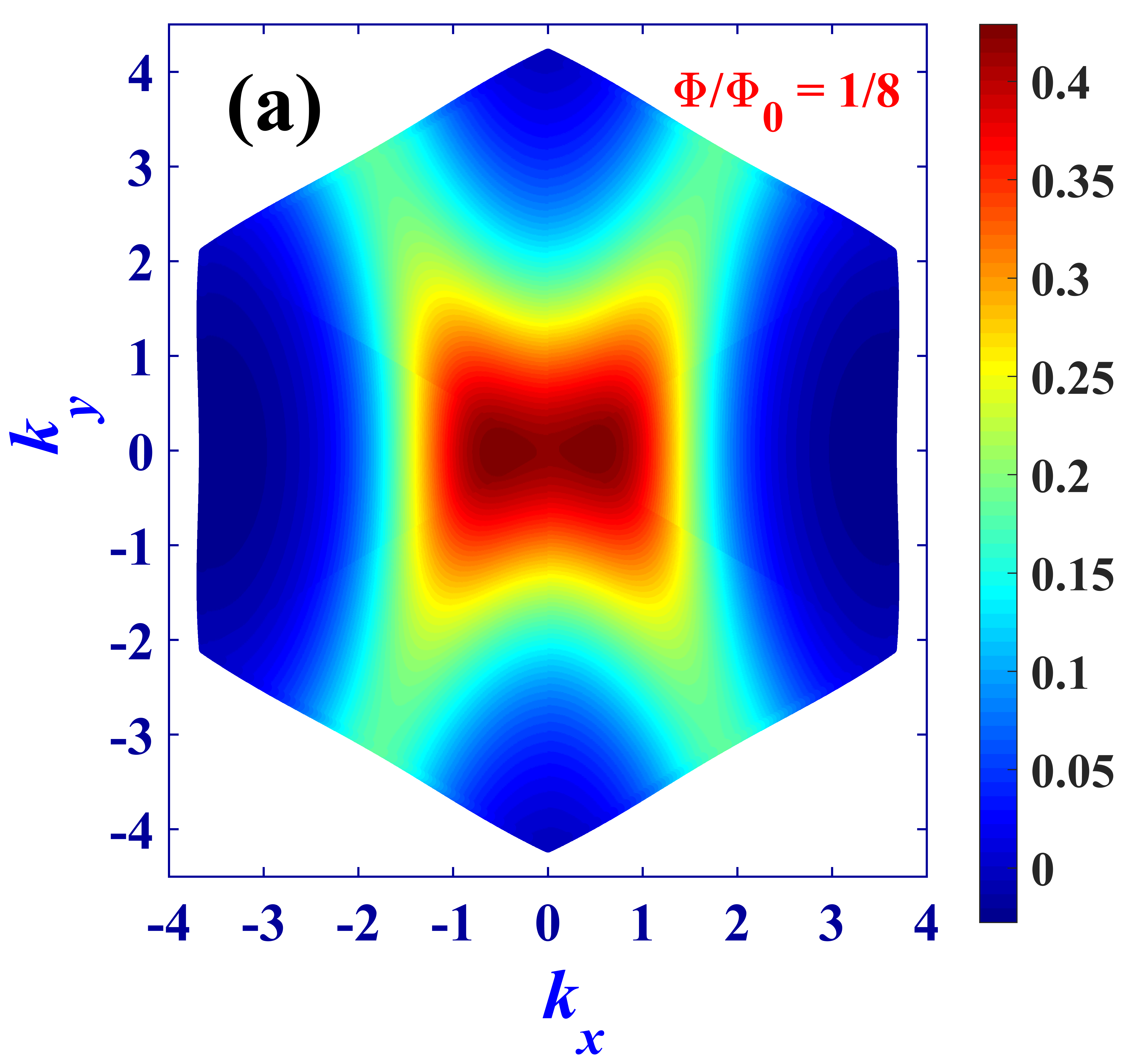}
\includegraphics[clip, width=0.55\columnwidth]{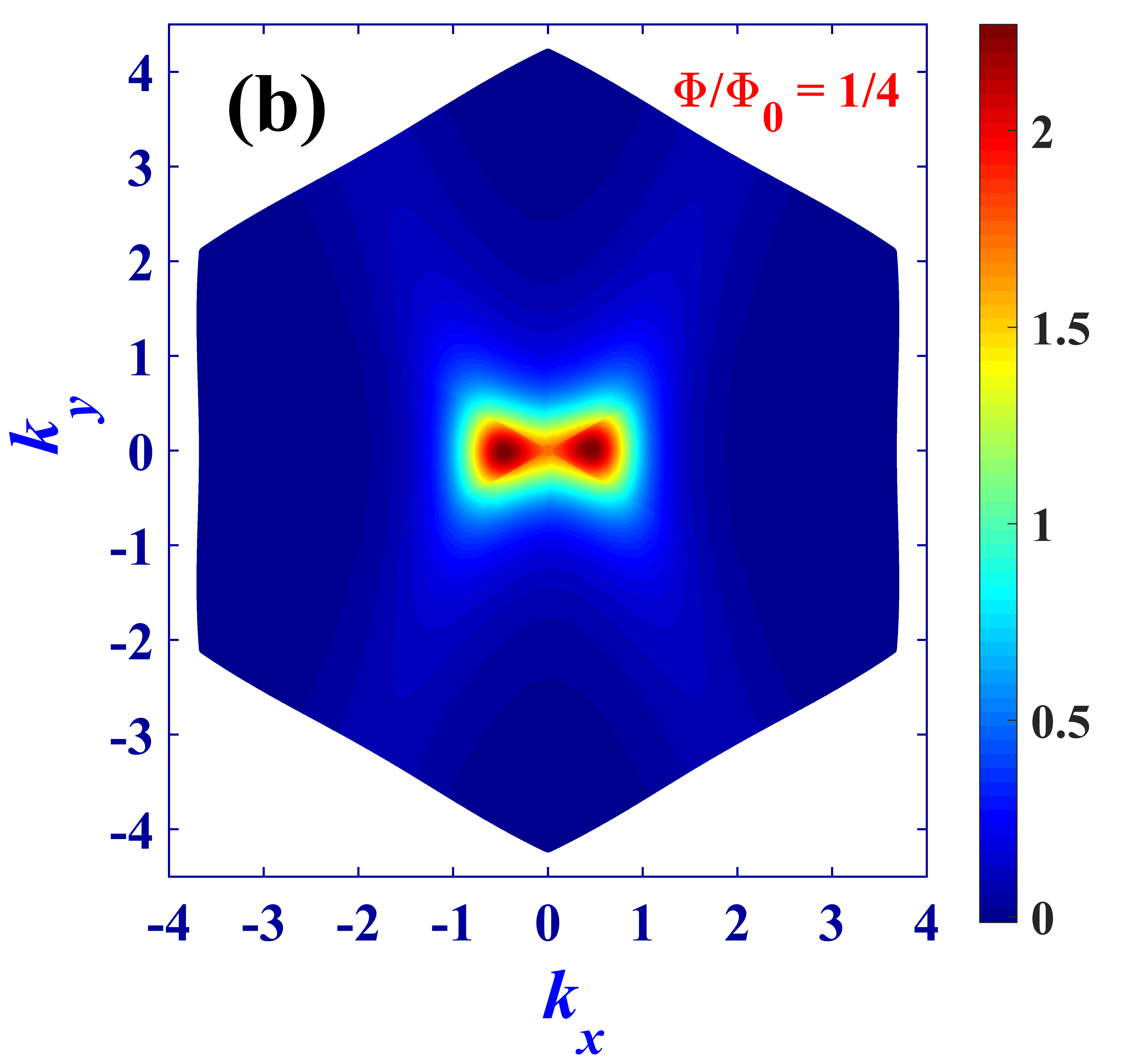}
\includegraphics[clip, width=0.55\columnwidth]{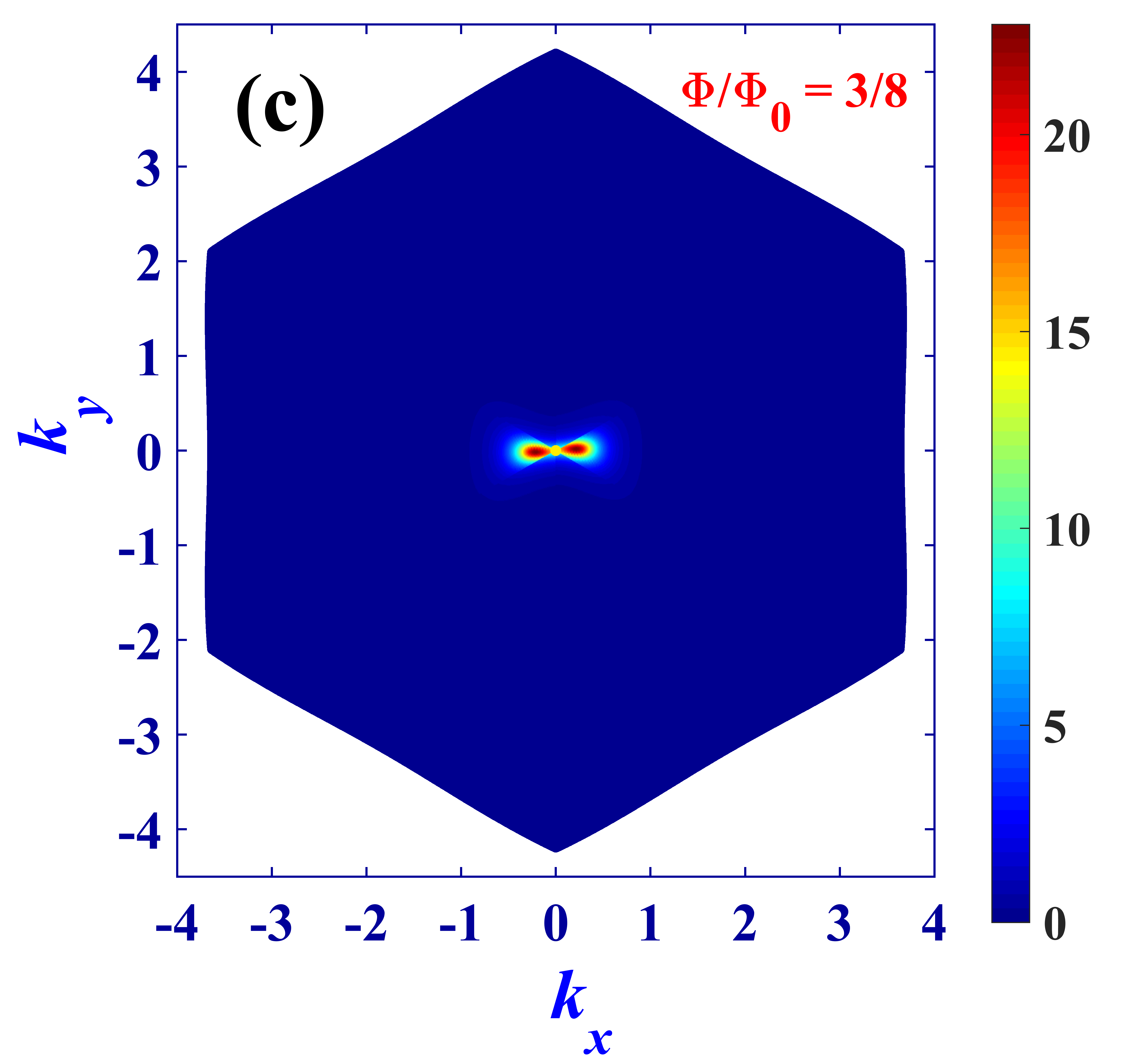}
\vskip 0.4cm
\includegraphics[clip, width=0.55\columnwidth]{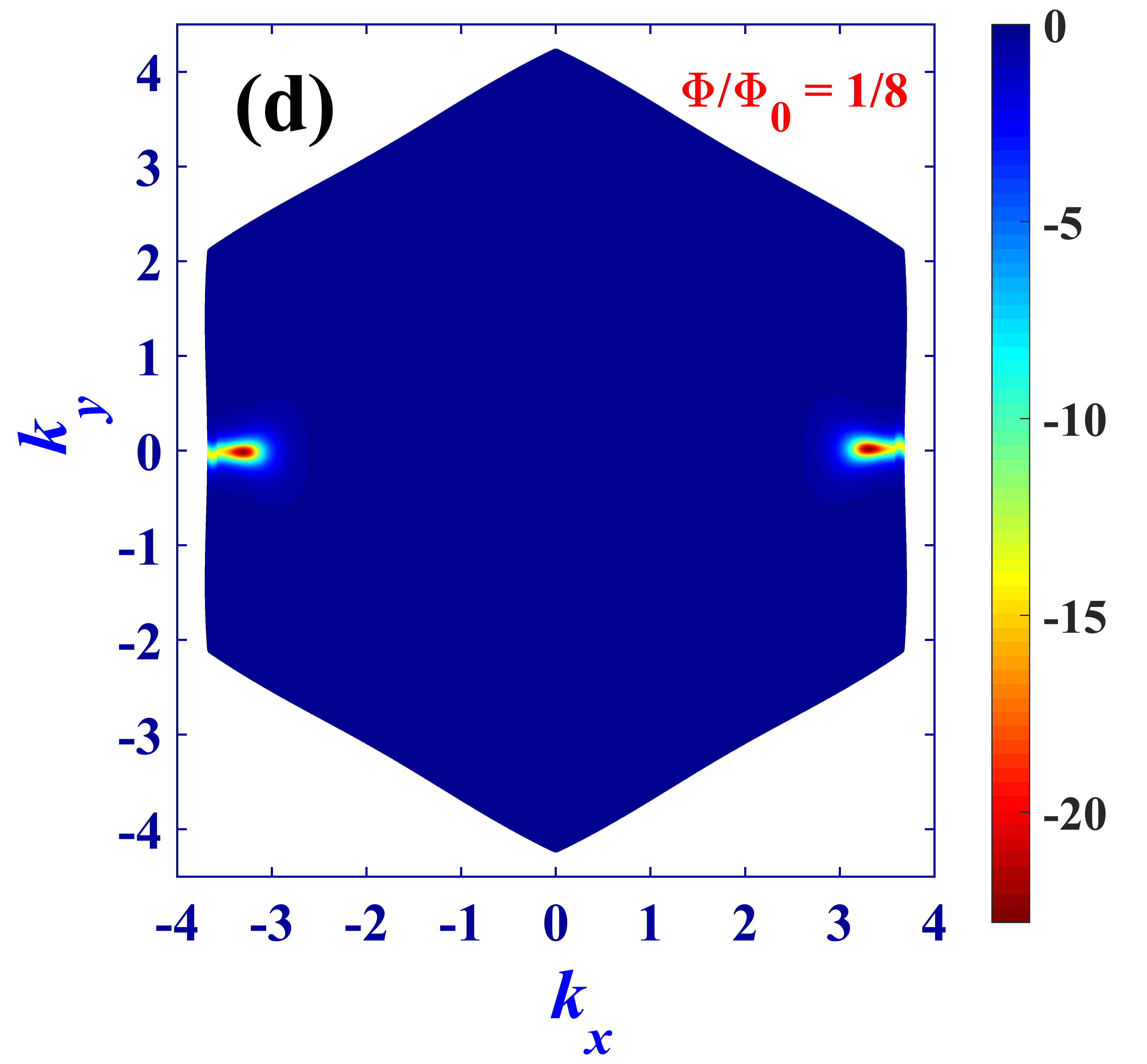}
\includegraphics[clip, width=0.55\columnwidth]{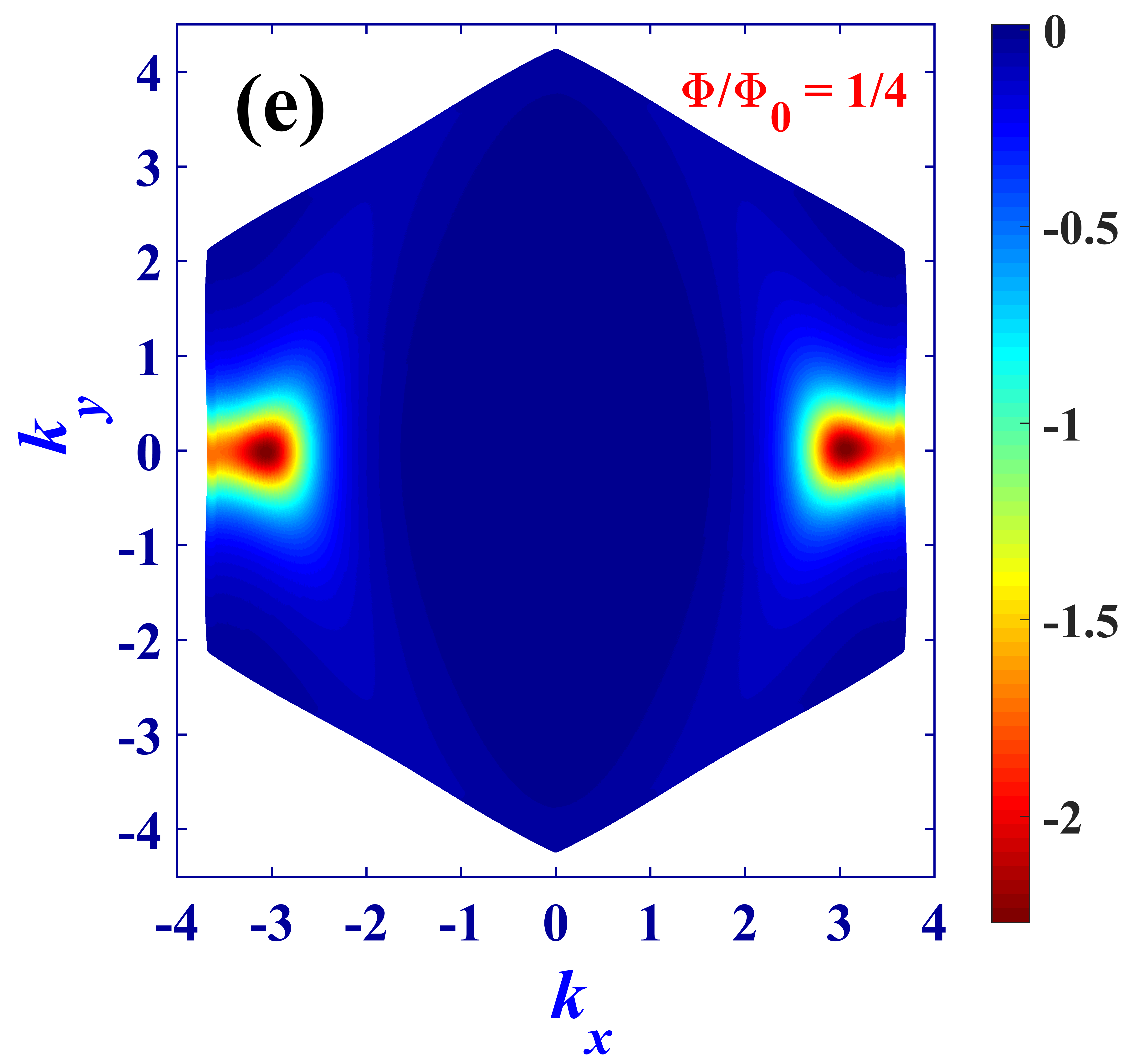}
\includegraphics[clip, width=0.55\columnwidth]{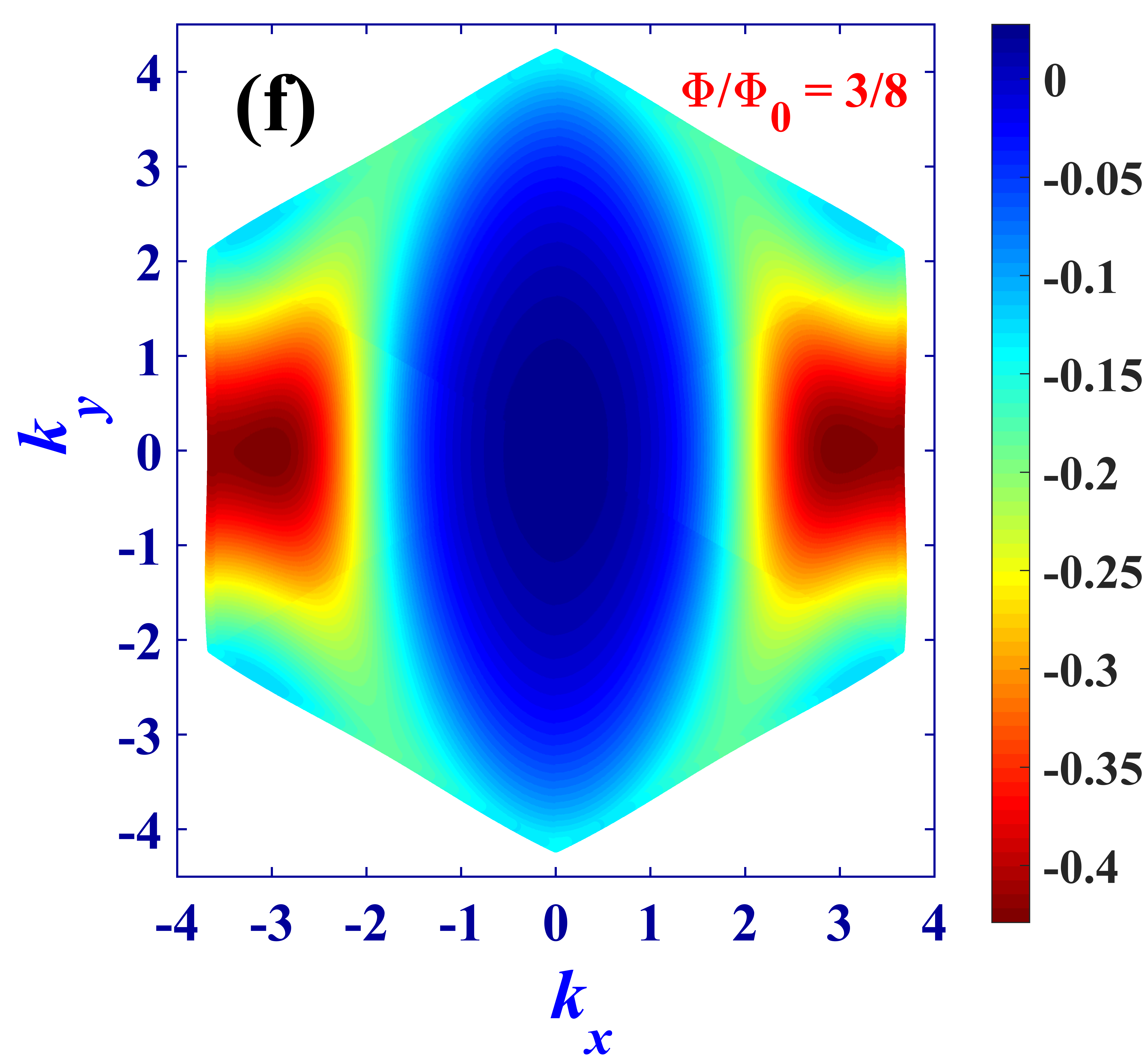}
\caption{The Berry curvatures corresponding to the nonzero integer values 
of the Chern numbers $\mathcal{C}_{1}=1$ (top panel) and 
$\mathcal{C}_{5}=-1$ (bottom panel) for three different special values of 
the magnetic flux $\Phi/\Phi_{0}= 1/8$, $1/4$, and $3/8$.}
\label{fig:BC}
\end{figure*}
We have already seen that, for zero flux, the bottom band is flat. We now switch 
on the magnetic flux and vary its amplitude. When we reach the special values 
$\Phi_{p}$, where $p = 1,2,3$ and $4$, it is found that the $(p+1)$-th band becomes 
flat. For these values of the flux, the FB energy is $E_{\emph{FB}_{2}} = -\sqrt{2}$, 
$E_{\emph{FB}_{3}} = 0$, $E_{\emph{FB}_{4}} = \sqrt{2}$, and $E_{\emph{FB}_{5}} = 2$ 
for respectively $p = 1, 2, 3$ and $4$. Such flux-controlled periodic generation of 
FBs is a very interesting feature and could be particularly of interest to address 
phenomena, such as flat band superconductivity or quantum Hall physics in this lattice. 

We have computed the CLS for each special value of the magnetic flux $\Phi=\Phi_{p}$. 
Finding the CLS with nonzero flux is a tedious task that has not been attempted in 
many occasions, apart from a very few rare instances~\cite{dias-prb-2011,dias-prb-2016}. 
The CLSs are depicted in Fig.~\ref{fig:CLS_with-flux}(a)-(d). 
We observe clusters spanning over 4 unit cells with nonzero values of the wave function 
amplitudes in various powers of $\alpha \equiv e^{i\theta}$, 
where $\theta = 2\pi\Phi/3\Phi_{0}$. 
Outside the finite cluster, the sites have strictly zero wave function amplitudes. 
It is to be noted that, the distribution of the wave function amplitudes are different 
for the CLSs, however those corresponding to $E = \pm \sqrt{2}$ are identical. 

In order to accurately find out if some bands are touching or are well separated 
from each other, we have computed the density of states (DOS) for each special value 
of the flux. The result is displayed in Fig.~\ref{fig:DOS}. The nature of the FB 
eigenstates, \ie, whether they are gapped or gapless with other dispersive bands 
in the system has also been identified in Fig.~\ref{fig:DOS}. It is found that, 
the band touching occurs only when the flux $\Phi/\Phi_{0} = p/2$ (where $p = 0$ or $1$), 
otherwise a large gap systematically separates the FB from the dispersive bands.
\section{Characterizing the topological states in the system}
\label{sec:topo-properties}
In this section, we present our findings on the topological properties appearing 
in the Stagome lattice model. Identifying the topological properties in FB systems has 
been active topic of interest to the community~\cite{Wen-PRL-2011,Das-Sarma-PRL-2011,
Neupert-PRL-2011,Pal-PRB-2018,Bhatta-Pal-PRB-2019} because of its relevance to quantum 
Hall physics and superconductivity in 2D lattice models. One of the most popular way 
to study the topological properties of a lattice model is to calculate the Berry 
curvature and the corresponding Chern numbers for each band appearing in the system.

Here, we follow this method, and calculate the Berry curvatures and Chern numbers using 
the following formulas~\cite{chen-jpcm2012,haldane-prl2004}: 
\begin{widetext}
\begin{equation}
\Omega_{n}(\bm{\vec{k}}) = 
\sum_{\mathcal{E}_{m} (\neq \mathcal{E}_{n})}\hspace{-3.5mm} 
\dfrac{-2\,\textrm{Im}\Big[ 
\langle \psi_{n}(\bm{\vec{k}}) | \dfrac{\partial \bm{\mathcal{H}}(\bm{\vec{k}})}{\partial k_{x}} | 
\psi_{m}(\bm{\vec{k}}) \rangle 
\langle \psi_{m}(\bm{\vec{k}}) | \dfrac{\partial \bm{\mathcal{H}}(\bm{\vec{k}})}{\partial k_{y}} | 
\psi_{n}(\bm{\vec{k}}) \rangle 
\Big]} 
{\big[\mathcal{E}_{n}(\bm{\vec{k}})-\mathcal{E}_{m}(\bm{\vec{k}})\big]^2},
\label{eq:BC}
\end{equation} 
\end{widetext}
where $\vert \psi_{n}(\bm{\vec{k}}) \rangle$ is the $n$-th band eigenstate of the 
Hamiltonian $\bm{\mathcal{H}}(\bm{\vec{k}})$ as defined in Eq.~\eqref{eq:hamilton-bloch} 
with energy $\mathcal{E}_{n}(\bm{\vec{k}})$. Using the expression in Eq.~\eqref{eq:BC}, 
one can easily compute the Chern number $\mathcal{C}_{n}$ associated with 
the `$n$'-th band as follows, 
\begin{equation}
\mathcal{C}_{n} = \frac{1}{2 \pi}\int_{BZ_{1}}
\Omega_{n} (\bm{\vec{k}})d{\bm{\vec{k}}},
\label{eq:Chern}
\end{equation}
where the integration (or summation in discrete case) is taken over the 
first (hexagonal) Brillouin zone $BZ_{1}$ of the Stagome lattice. 

It is clear that if there is a degeneracy in the energy eigenvalues between any 
two bands, then $\Omega_{n} (\bm{\vec{k}})$ diverges. So, the first and foremost 
thing we need to have in our system is band gaps between each band. In our model, 
this condition is accomplished by the introduction of the magnetic flux which 
acts as a perturbation to gap out all the bands in the system. Under this condition, 
we obtain nonzero integer values of the Chern numbers, \viz, $\mathcal{C}_{1}=1$ 
for the lowest band ($n=1$) and $\mathcal{C}_{5}=-1$ for the top band ($n=5$) for 
flux values in the regime $0<\Phi<\Phi_{0}/2$ as shown in Fig.~\ref{fig:Chern-number}. 
We observe that, the three other bands have a vanishing Chern number, 
and thus they are topologically trivial. Note that, for values of the 
flux $\Phi=p\times \Phi_{0}/2$, $\mathcal{C}_{1}$ and $\mathcal{C}_{5}$ reverse their 
sign (see Fig.~\ref{fig:Chern-number}). From this, we may conclude that $\Phi = \Phi_{0}/2$ 
acts as a critical point, where there is a topological phase transition occurring 
in the Stagome lattice. This is due to the fact that, there is a band-touching and band 
gap reopening happen at $\Phi = \Phi_{0}/2$. 
It is to be noted, a perfectly FB occurring with short-range hopping cannot 
have a finite Chern number. In contrast, nearly FB can have non vanishing Chern 
numbers~\cite{Wen-PRL-2011,Das-Sarma-PRL-2011,Neupert-PRL-2011,Pal-PRB-2018,Bhatta-Pal-PRB-2019}. 
In the case of the Stagome lattice, we identify similar feature near $\Phi=0$ and $\Phi=\Phi_{0}/2$.

We now display the distribution of the Berry curvatures (BC) for the two topologically 
nontrivial Chern bands ($n=1$ and $n=5$) and for the three special values of the 
magnetic flux, \viz, $\Phi= \Phi_{0}/8$, $\Phi_{0}/4$, and $3\Phi_{0}/8$, respectively 
in Fig.~\ref{fig:BC}. First, for any values of the flux, the BC weight is the found to 
be the largest in the vicinity of the $\Gamma$-point (respectively $M$-point) of the 
Brillouin zone for the first (respectively $5$th) band. More specifically, 
for $\Phi= \Phi_{0}/8$, the BC is extremely localized in the vicinity of the $M$-point 
for the $5$th band and much more spread out near the $\Gamma$-point for the first band. 
For $\Phi= 3\Phi_{0}/8$, the situation is reversed. Finally, for $\Phi = \Phi_{0}/4$, 
the structure and intensity are identical, the difference being that one is located at 
the $\Gamma$-point, the other at the $M$-point. The origin of the symmetries in the 
spectrum and in the Berry curvatures are clarified in the Appendix~\ref{appendix-2}.
%
\begin{figure}[ht]
\centering
\includegraphics[clip, width=\columnwidth]{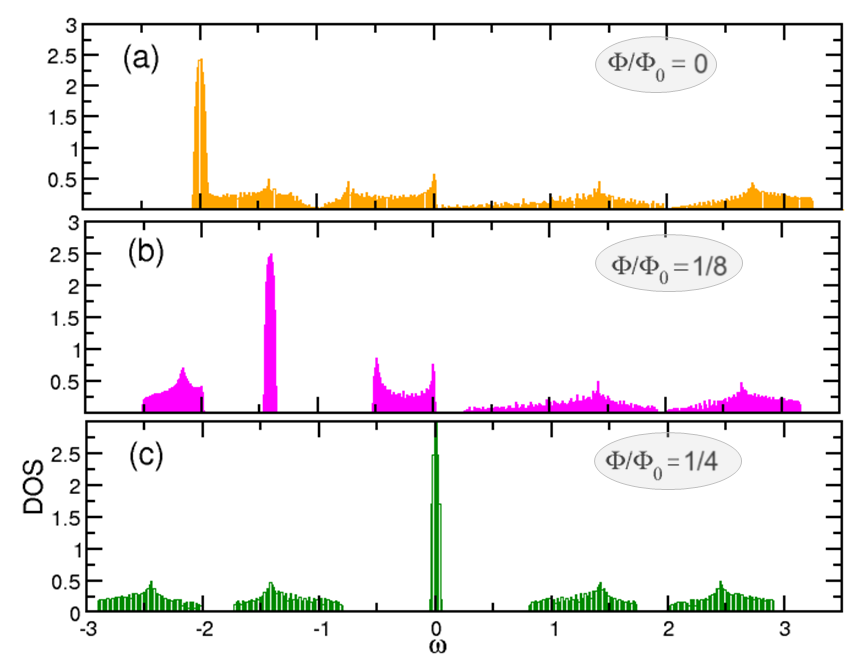}
\caption{Density of states as a function of the energy ($\omega$) in the disordered 
Stagome lattice for (a) $\Phi/\Phi_{0} = 0$, (b) $1/8$ and (c) $1/4$.
The system contains $48\times48$ unit cells and the average is performed over 
10 configurations of the off-diagonal disorder.}
\label{fig:dos-disorder}
\end{figure}
%
\begin{figure}[ht]
\centering
\includegraphics[clip, width=\columnwidth]{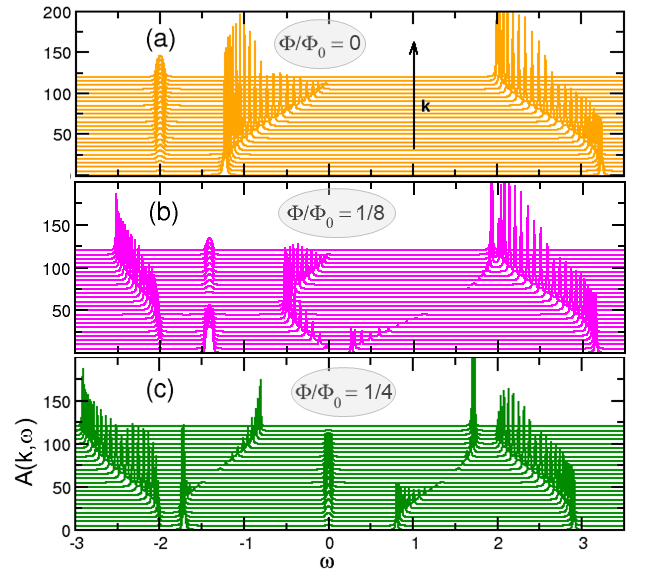}
\caption{Average Spectral function $A(\textbf{k},\omega)$ as a function of 
$\omega$ where $\textbf{k}$ is chosen along the $y$-direction in the Brillouin zone. 
For the sake of visibility a constant $C_{\textbf{k}}= \alpha|\textbf{k}|$ is added to
$A(\textbf{k},\omega)$ (here $\alpha \approx 20)$. In panel (a), (b) and (c) 
respectively the flux is $\Phi/\Phi_{0} = 0$, $1/8$ and $1/4$. The system contains 
$48\times48$ unit cells and the average is performed over 10 different configurations 
of the off-diagonal disorder.}
\label{fig:sqw-disorder}
\end{figure}
%
\section{Effects of the disorder}
\label{sec:disorder}
In this section, our purpose is to address the robustness of our results when disorder is 
present in the system. Here, we consider that the disorder is of an off-diagonal type which 
implies the real-space diagonalization of the Hamiltonian and a systematic average the 
observables over several configurations of the disorder. Off-diagonal disorder means that 
for a given pair of nearest-neighbor sites $(i\alpha,j\beta)$, the hopping term 
$t_{i\alpha,j\beta}$ in the clean system becomes $t_{i\alpha,j\beta}(1+\delta w_{i\alpha,j\beta})$ 
where $\delta w_{i\alpha,j\beta}$ is randomly chosen in the interval $[-W/2,W/2]$. In this section, 
we set $W = 0.1$ and consider a large system that contains $N_{c} = 48 \times 48$ cells. 

In Fig.~\ref{fig:dos-disorder}, the averaged density of states (over the disorder) is depicted as 
a function of the energy for $\Phi/\Phi_{0} = 0$, $1/8$ and $1/4$. As can be seen, the data 
are very similar to that plotted in Fig.~\ref{fig:DOS} in the clean system. The main difference 
concerns the flat band, which is still located at the same energy for each value of the flux 
but now has a finite width equal to $W$. 

The DOS is not sufficient to fully estimate the impact of the disorder. It is therefore 
interesting to also calculate the spectral function which, in the disordered system, is defined by,
\begin{equation}
A(\textbf{k},\omega) = -\frac{1}{\pi} \sum_{\alpha} 
\Im (\langle G_{\alpha\alpha}^{c}(\textbf{k},\omega)\rangle_{dis}),
\label{eq:sqw}
\end{equation}
where $\langle... \rangle_{dis}$ means average over the configurations of the disorder.
$G_{\alpha\alpha}^{c}(\textbf{k},\omega)=\frac{1}{N_{c}} \sum_{i,j}e^{i\textbf{k}. 
(\textbf{R}_{i\alpha}-\textbf{R}_{j\alpha})} G_{i\alpha,j\alpha}^{c}$,
$i,j$ are cell indices, $\alpha$ is the orbital index and 
$G_{i\alpha,j\alpha}^{c} = \langle i\alpha \vert\frac{1}{(\omega + i\eta) 
\hat{I} - \hat{H}} \vert j\alpha\rangle$, where $\hat{I}$ is the matrix identity 
of size $5\,N_c \times 5\,N_c $, and $i\eta$ is a very small imaginary part ($\eta \ll W$). 
The subscript `$c$' corresponds to a given configuration of the disorder. 

In Fig.~\ref{fig:sqw-disorder}, the spectral function is depicted in the disordered Stagome 
lattice where $\textbf{k}$ is chosen along the $y$-direction in the Brillouin zone.
Let us focus our attention on the quasi FB. First, for each pair $(\Phi,\textbf{k})$, we 
observe that the peak is exactly located at the same position as in the clean system, 
namely $-2$, $-\sqrt{2}$ and $0$, respectively for
$\Phi/\Phi_{0} = 0$, $1/8$ and $1/4$. In addition, for a given $\textbf{k}$ the width 
of the quasi FB peak is momentum-independent, $\Delta \omega_{\textbf{k}} = 0.1$. 
We notice that same features have been observed for other directions in the Brillouin zone.
These findings imply that the concept of flat-band is still meaningful even in the presence 
of disorder and that the physics reported in the clean system is robust against random 
fluctuations in the hoppings. It is to be noted that, we have considered as well the case of 
diagonal disorder (random on-site potentials) and could reach the same conclusions. 
\vskip 0.8cm
\section{Conclusion}
\label{sec:conclu}
To conclude, we have studied the intriguing flat band physics in a newly proposed 
two-dimensional lattice geometry named as Stagome lattice, in which one can systematically 
make any of the bands completely flat by tuning an external magnetic flux. We have 
systematically calculated the CLS, the Berry curvatures and the corresponding Chern numbers. 
It is found that, some of the bands are topologically nontrivial and our model exhibits a 
topological phase transition. We have as well shown that our findings are 
robust against fluctuations due to presence of a small amount of disorder. 
With the recent progress in artificial lattices, we believe that, the flat-band localization 
could be explored in the photonic Stagome lattice, that could be fabricated using laser 
writing technology. Furthermore, the CLS being extended over several unit cells, and with 
the freedom to tune the position of the FB, the Stagome lattice could be a promising 
candidate to explore the FB superconductivity in a controllable fashion.
\begin{acknowledgments}
BP gratefully acknowledges the funding from CNRS through CPTGA visiting faculty 
scheme hosted at the N\'{e}el Institute, where this work was initiated. The authors 
are thankful to Maxime Thumin for useful discussions during the initial stage of this work.  
\end{acknowledgments}
%
\renewcommand\thefigure{A.\arabic{figure}}    
\setcounter{figure}{0} 
\begin{figure*}[ht]
\centering
\includegraphics[clip, width=0.4\columnwidth]{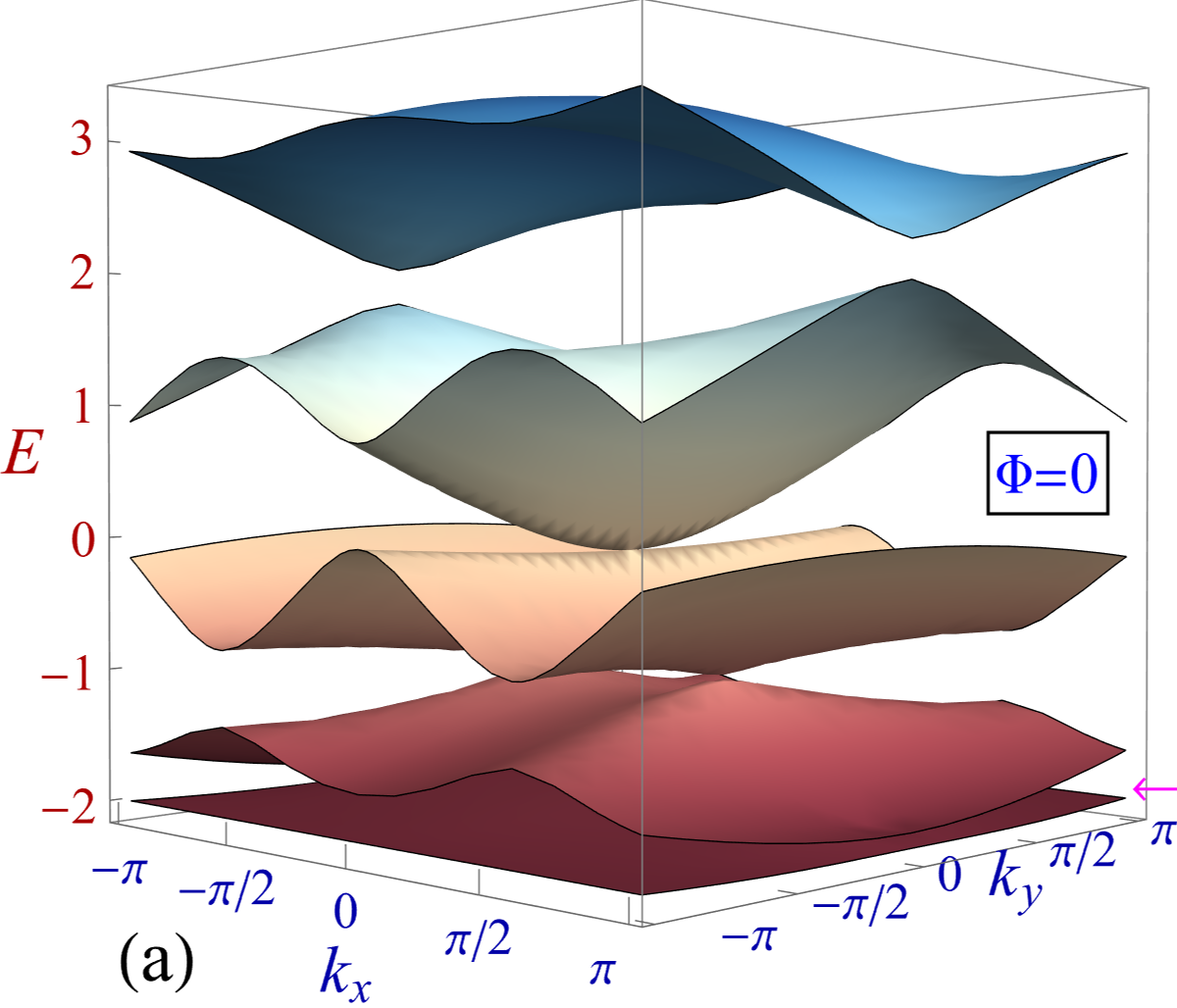}
\includegraphics[clip, width=0.4\columnwidth]{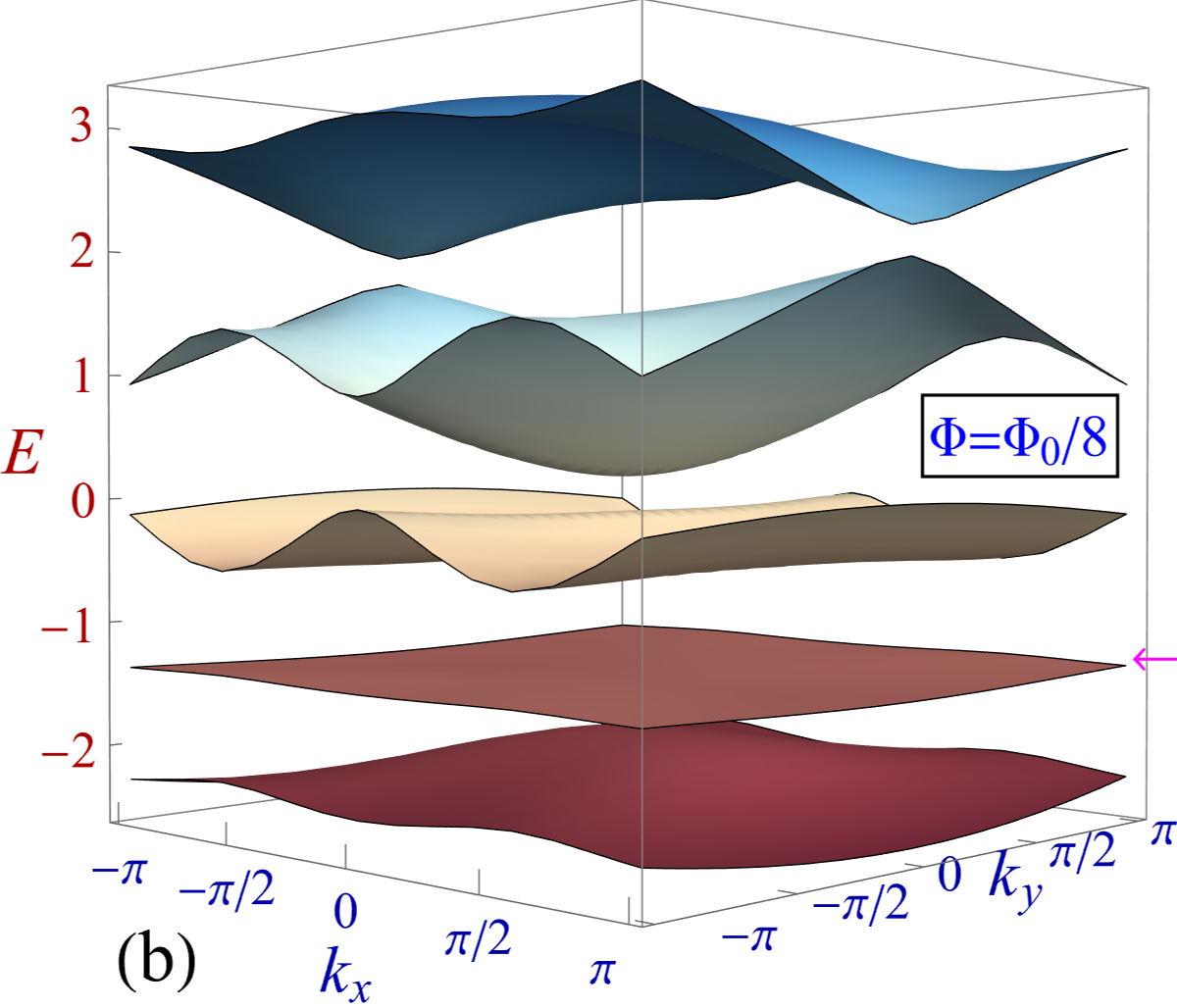}
\includegraphics[clip, width=0.4\columnwidth]{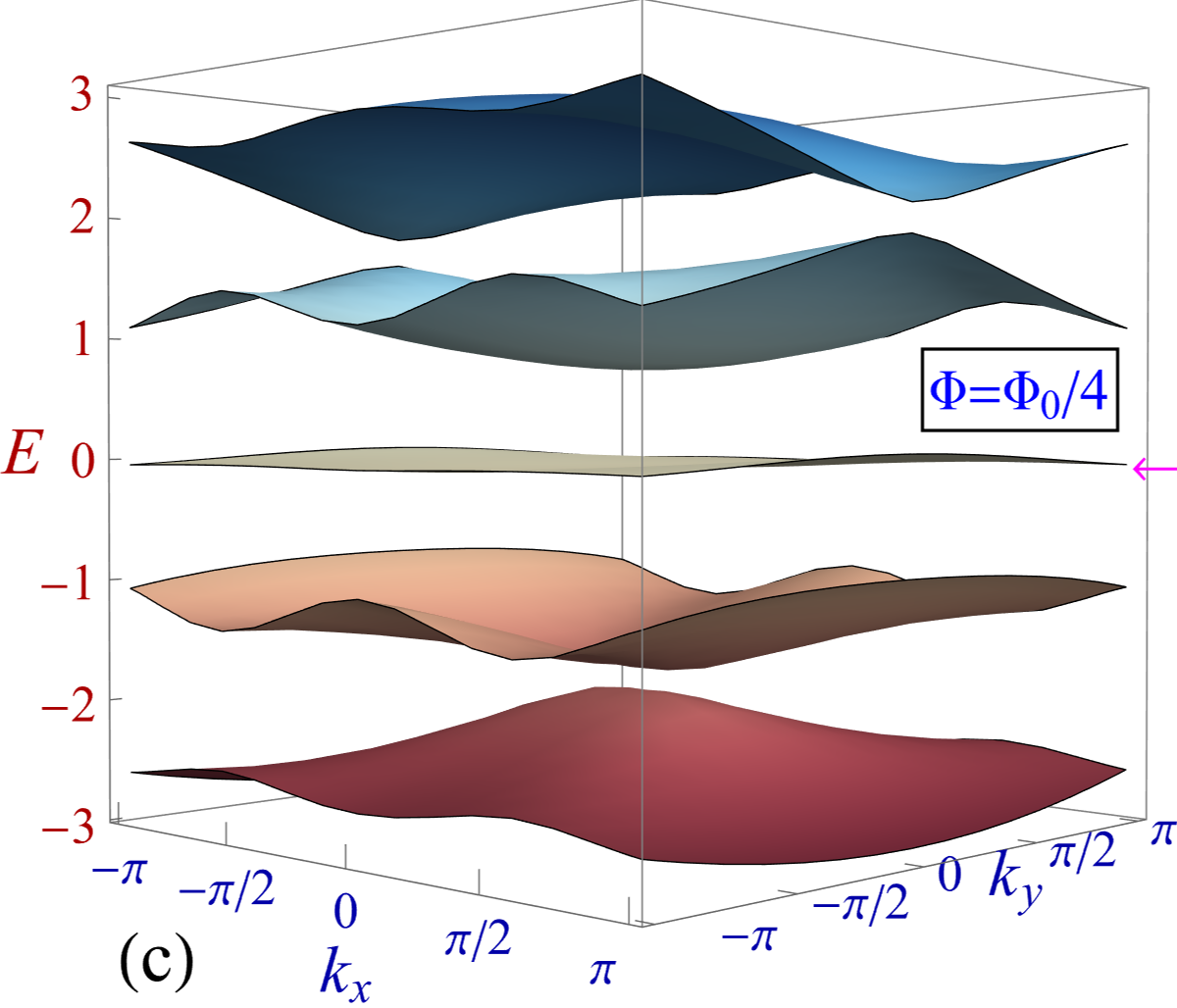}
\includegraphics[clip, width=0.4\columnwidth]{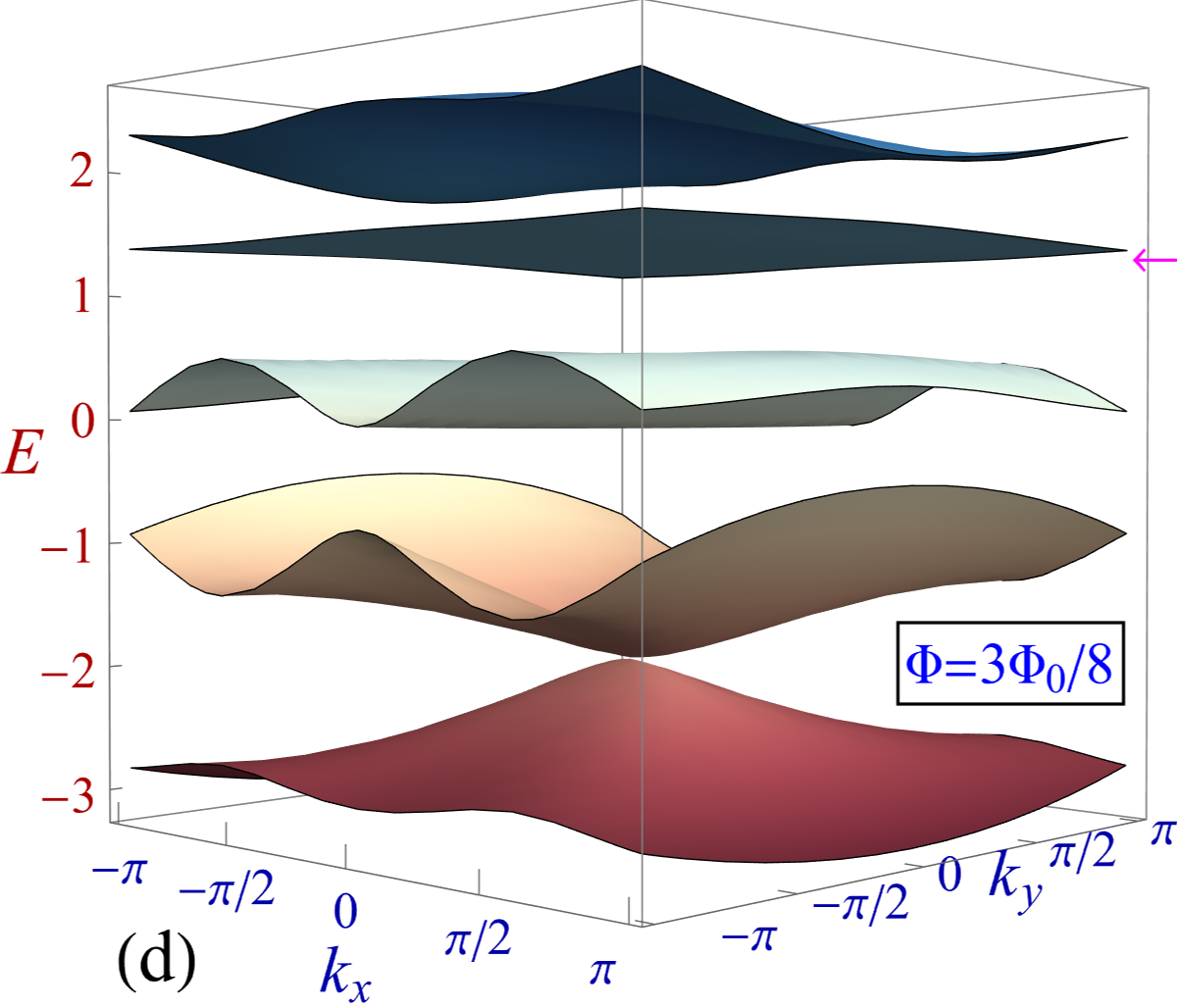}
\includegraphics[clip, width=0.4\columnwidth]{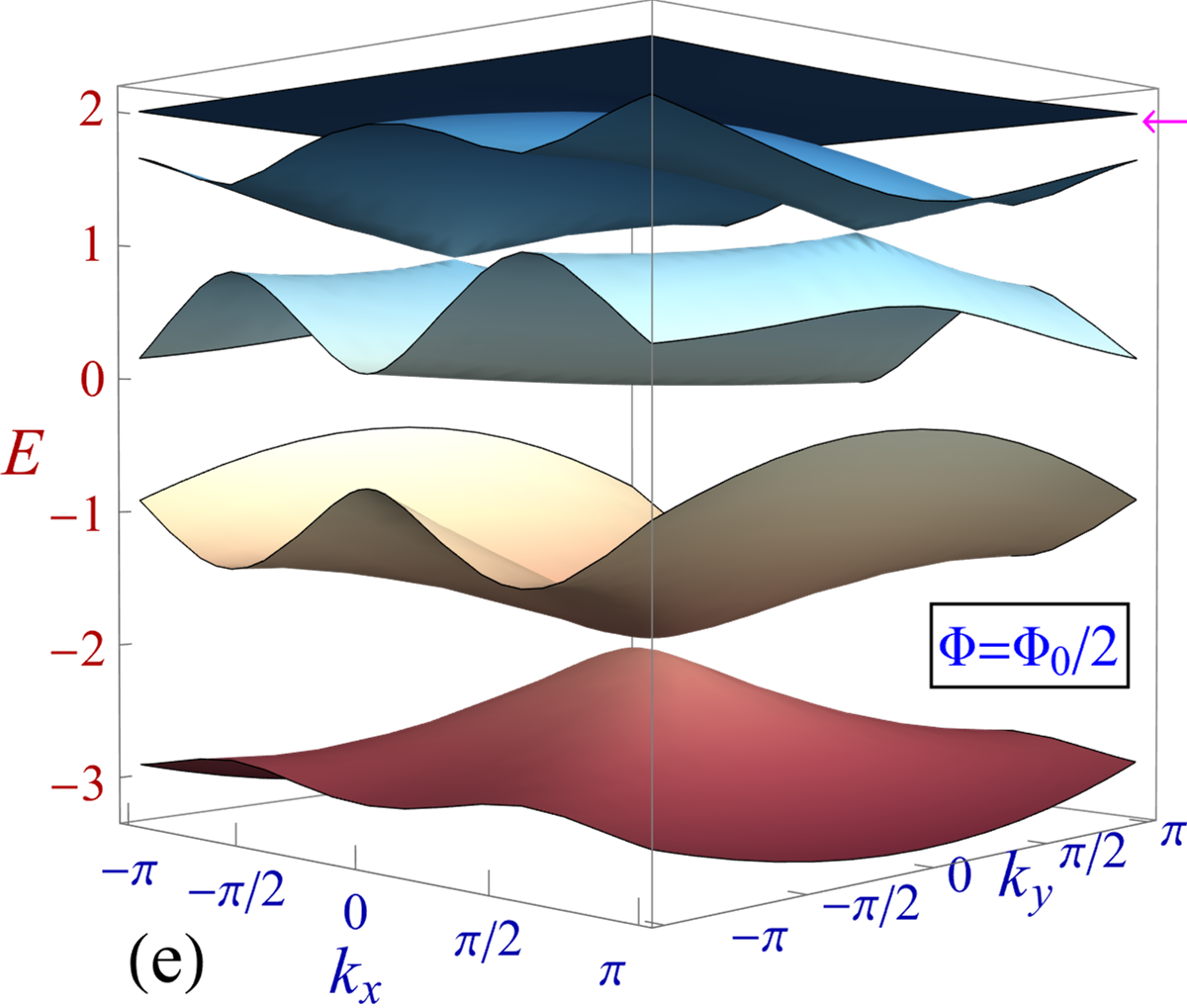}
\caption{The electronic band structure of the Stagome lattice model 
with next-nearest neighbor hopping $t^{\prime}=0.1t$. The other 
parameters remain same as in Fig.~\ref{fig:Band-n-CLS_without-flux} and 
Fig.~\ref{fig:Band-structures_with-flux}. The quasi-flat bands are 
marked with arrows in the plots.}
\label{fig:Band-structures_with-NNN}
\end{figure*}
%
\appendix
\section{Effect of the next-nearest neighbor hopping on the flat bands}
\label{appendix-1}
With inclusion of the next-nearest neighbor hopping term $t^{\prime}$ in the Stagome 
lattice model, the Hamiltonian in Eq.~\eqref{eq:hamilton-bloch} is modified 
as $\bm{\mathcal{H}}+\bm{\Delta\mathcal{H}}$, where
\begin{align}
\bm{\Delta\mathcal{H}} =
t^{\prime}\left(\def\arraystretch{1.0} 
\begin{matrix}
0  &  0  &  0  &  1  &  0 \\
0  &  0  &  0  &  0  &  1 \\
0  &  0  &  0  &  0  &  0 \\
1  &  0  &  0  &  0  &  0 \\
0  &  1  &  0  &  0  &  0  \\
\end{matrix}\right).
\label{eq:hamilton-bloch-with-NNN}
\end{align}
For a small value of $t^{\prime}=0.1t$, the completely flat bands shown in 
Fig.~\ref{fig:Band-n-CLS_without-flux} and 
Fig.~\ref{fig:Band-structures_with-flux} become \emph{nearly flat}. 
This is displayed in Fig.~\ref{fig:Band-structures_with-NNN}. 
The bandwidth of theses quasi-flat bands is $\sim t^{\prime}$. The values 
of the corresponding Chern numbers under this condition also remain the same as 
depicted in Fig.~\ref{fig:Chern-number}. It is to be noted that, if we consider 
large values of $t^{\prime}$, then all these features disappear. 
\\
\section{Symmetries of the Hamiltonian}
\label{appendix-2}
In this appendix, our purpose is to shed light on the origin of the symmetries observed 
in the dispersions and in the Berry curvatures. It is convenient to include 
the variables $t$, $\lambda$, and $\theta$ explicitly in the definition of the Hamiltonian, 
where $t$ and $\lambda$ are assumed to be real. 
\\ 
We write,
\begin{align}
\bm{\mathcal{H}}(t,\lambda,\theta,\bm{\vec{k}}) =
\left(\def\arraystretch{1.5} \begin{matrix}
0 &  te^{i\theta}  &  te^{-i\theta}  &  0  
&  \lambda e^{-i\bm{\vec{k}}\cdot\bm{\vec{r}_{2}}} \\
te^{-i\theta}  &  0  &  te^{i\theta}  &  
\lambda e^{i\bm{\vec{k}}\cdot\bm{\vec{r}_{1}}}  &  0 \\
te^{i\theta}  &  te^{-i\theta}  &  0 &  
te^{-i\theta}  &  te^{i\theta}  \\
0  &  \lambda e^{-i\bm{\vec{k}}\cdot\bm{\vec{r}_{1}}}  &  
te^{i\theta}  &  0  &  te^{-i\theta}  \\
\lambda e^{i\bm{\vec{k}}\cdot\bm{\vec{r}_{2}}}  &  
0  &  te^{-i\theta}  &  te^{i\theta}  &  0  \\
\end{matrix}\right)
\label{eq1}
\end{align}
In this appendix we show that, the Hamiltonian is invariant under the transformation:
$\theta \rightarrow -\theta^{\prime} + \frac{\pi}{3}$, where we recall that 
$\theta = \frac{2\pi\Phi}{3\Phi_0}$.
\\
First, one has the trivial property,
\begin{align}
\bm{\mathcal{H}}(t,\lambda,-\theta,-\bm{\vec{k}})=
\bm{\mathcal{H}}^{*}(t,\lambda,\theta,\bm{\vec{k}})
\label{eq2}
\end{align}
For any given $\bm{\vec{k}}$ in the Brillouin zone, there is a corresponding $\bm{\vec{k}}^{\prime}$, 
such that $\bm{\vec{k}}^{\prime}\cdot\bm{\vec{r}}_i=\bm{\vec{k}}\cdot\bm{\vec{r}}_i +\pi$ where $i=1,2$.
As a consequence one obtains the relation,
\begin{align}
\bm{\mathcal{H}}(t,-\lambda,\theta,\bm{\vec{k}})=
\bm{\mathcal{H}}(t,\lambda,\theta,\bm{\vec{k}}^{\prime})
\label{eq3}
\end{align}
We recall that the Hamiltonian is invariant under the transformation $\frac{\Phi}{\Phi_0} 
\rightarrow \frac{\Phi}{\Phi_0} \pm 1$, which corresponds to
$\theta \rightarrow \theta \pm \dfrac{4\pi}{3}$.
As a consequence one finds that,
\begin{align}
\bm{\mathcal{H}}(t,\lambda,\theta \pm \frac{4\pi}{3},\bm{\vec{k}})=
\bm{\mathcal{H}}(t,\lambda,\theta,\bm{\vec{k}})
\label{eq4}
\end{align}
We now combine these set of equations Eq.~\eqref{eq1} to Eq.~\eqref{eq4} to reach our conclusion: 
\begin{align*}
& \bm{\mathcal{H}}(t,\lambda,\theta',\bm{\vec{k}}) \xrightarrow{\eqref{eq4}}
\bm{\mathcal{H}}(t,\lambda,-\theta-\pi,\bm{\vec{k}}) \xrightarrow{\eqref{eq2}}\\
& \bm{\mathcal{H}}^{\ast}(t,\lambda,\theta+\pi,-\bm{\vec{k}}) \xrightarrow{\eqref{eq1}}
\bm{\mathcal{H}}^{\ast}(-t,\lambda,\theta,-\bm{\vec{k}}) \xrightarrow{\eqref{eq1}}\\ 
& -\bm{\mathcal{H}}^{\ast}(t,-\lambda,\theta,-\bm{\vec{k}}) \xrightarrow{\eqref{eq3}}
-\bm{\mathcal{H}}^{\ast}(t,\lambda,\theta,-\bm{\vec{k}}') 
\end{align*}
Finally, we end up with the final symmetry relation,
\begin{align}
\bm{\mathcal{H}}(t,\lambda,\theta',\bm{\vec{k}})=
-\bm{\mathcal{H}}^{*}(t,\lambda,\theta,-\bm{\vec{k}}')
\label{eq5}
\end{align}
This equation explains all the symmetries observed in (i) the dispersions, (ii) the DOS, 
and (iii) the Berry curvatures. It nicely clarifies as well why the Berry curvature are 
identical and centered around the $\Gamma$-point for band 1 
when $\theta = \pi/12$ (Fig.~\ref{fig:BC}{a}) and around the M-point for band 5 
when $\theta^{\prime}= -\theta + \pi/3 = \pi/4$.

\end{document}